\documentclass[journal,letterpaper,twocolumn]{IEEEtran}
\usepackage{cite}
\usepackage[cmex10]{amsmath}
\usepackage{tabularx,colortbl}
\definecolor{Gray}{gray}{0.9}

\usepackage{amsmath,amssymb,amsthm,mathrsfs,dsfont}

\usepackage{amsmath}
\usepackage{amssymb}

\usepackage{booktabs}
\usepackage{threeparttable}
\usepackage{multirow}

\usepackage{algorithm}
\usepackage{algpseudocode}

\usepackage{subfigure}
\usepackage{stfloats}

\usepackage[utf8]{inputenc}

\newtheorem{lemma}{Lemma}
\newtheorem{proposition}{Proposition}

\IEEEoverridecommandlockouts
\ifCLASSINFOpdf
  \usepackage[pdftex]{graphicx}
  \DeclareGraphicsExtensions{.pdf,.jpeg,.png}
\else
  \usepackage[dvips]{graphicx}
  \DeclareGraphicsExtensions{.eps}
\fi

\begin{document}

\title{Distributed and Optimal Resource Allocation for Power Beacon-Assisted Wireless-Powered Communications}

\author{\IEEEauthorblockN{Yuanye Ma, \textit{Student Member, IEEE}, He (Henry) Chen, \textit{Student Member, IEEE}, Zihuai Lin, \textit{Senior Member, IEEE}, Yonghui Li, \textit{Senior Member, IEEE}, and Branka Vucetic, \textit{Fellow, IEEE}\\}

\thanks{This work of Yuanye Ma was supported by China Scholarship Council and Norman I Price Supplementary Scholarship. The work of He Chen was supported by International Postgraduate Research Scholarship (IPRS) and Australian Postgraduate Award (APA).}
\thanks{Part of this work was presented at the IEEE International Conference on Communications (ICC), London, UK, June 2015~\cite{YuanyeICC}.}
\thanks{The authors are with the School of Electrical and Information Engineering, the University of Sydney, NSW 2006, Australia (e-mail: \{yuanye.ma, he.chen, zihuai.lin, yonghui.li, branka.vucetic\}@sydney.edu.au).}
}

\maketitle
\begin{abstract}

In this paper, we investigate optimal resource allocation in a power beacon-assisted wireless-powered communication network (PB-WPCN), which consists of a set of hybrid access point (AP)-source pairs and a power beacon (PB). Each source, which has no embedded power supply, first harvests energy from its associated AP and/or the PB in the downlink (DL) and then uses the harvested energy to transmit information to its AP in the uplink (UL). We consider both cooperative and non-cooperative scenarios based on whether the PB is cooperative with the APs or not. For the cooperative scenario, we formulate a social welfare maximization problem to maximize the weighted sum-throughput of all AP-source pairs, which is subsequently solved by a water-filling based distributed algorithm. In the non-cooperative scenario, all the APs and the PB are assumed to be rational and self-interested such that incentives from each AP are needed for the PB to provide wireless charging service. We then formulate an auction game and propose an auction based distributed algorithm by considering the PB as the auctioneer and the APs as the bidders. Finally, numerical results are performed to validate the convergence of both the proposed algorithms and demonstrate the impacts of various system parameters.


\end{abstract}

\IEEEpeerreviewmaketitle

\begin{IEEEkeywords}
Power beacon-assisted wireless-powered communication network (PB-WPCN), wireless energy transfer (WET), resource allocation, auction theory.
\end{IEEEkeywords}

\section{Introduction}

Recently, wireless energy transfer (WET) technologies have drawn wide attention with their capability of energy supply \cite{zhangruimaga}. Conventionally, the very limited energy of wireless devices powered by batteries largely constrains the communication performance in many practical cases such as wireless sensor networks \cite{nishimoto2010prototype}. Also, the battery replacement for wireless devices is not always convenient or feasible in many applications. These challenging issues have boosted the development of WET technologies, which enable wireless devices to harvest energy from wireless signals for their intended applications. As one category of existing WET techniques, the radio frequency (RF)-enabled WET \cite{lu2014wireless} considered in this paper, provides the feasibility of a long-range energy transfer (up to tens of meters \cite{Cuttinglastwire}) compared to other technologies, such as inductive coupling \cite{want2006introduction} and magnetic resonance coupling~\cite{kurs2007wireless}. RF-enabled WET has not been widely used in practice largely due to the high propagation loss of RF signals. However, due to the latest breakthroughs in wireless communications, namely small cells \cite{andrews2014will}, transmission using large-scale antenna arrays (i.e., massive MIMO) \cite{larsson2013massive}, millimeter-wave communications \cite{wang2015multi}, the transmission distances would be dramatically reduced and sharp beamforming could be enabled, which will significantly reduce the propagation loss and achieve much higher WET efficiencies \cite{Cuttinglastwire}. Furthermore, the energy consumption of users will be continuously reduced by the advancements in low-power electronics \cite{6697937}. Thus, we believe that the RF-enabled WET has a great potential to be widely implemented in the next-generation wireless communication systems. Then, by considering the RF-enabled WET technique, a fully wireless-powered communication network (WPCN) can be established with no need of battery replacement \cite{zhangruimaga}. In a WPCN, wireless devices are only powered by WET in the downlink (DL) and transmit their information using the harvested energy in the uplink (UL)\cite{6678102}.

In open literature, there have been several papers that focused on the design of WPCNs for different setups \cite{6678102,30,28,29,6697937}. In \cite{6678102}, a ``harvest-then-transmit" protocol was proposed for a multi-user WPCN, where users first harvest energy from RF signals broadcast by a single antenna hybrid access point (AP) in the DL and then transmit information to the AP in the UL via time division multiple access (TDMA). Moreover, the DL WET time from the AP and UL information transmission time of individual users were jointly optimized to maximize the system sum-throughput. \cite{30} extended \cite{6678102} to a multi-antenna WPCN scenario, where a multi-antenna AP enables the simultaneous UL transmission via space division multiple access (SDMA). \cite{28} and \cite{29} considered full-duplex WPCNs, where a full-duplex AP is adopted to provide the simultaneous DL WET and UL information transmission. Moreover, \cite{Chen_TSP_2015_Harvest,Ju_Globecom_2014,Gu_ICC_2015} focused on the development of cooperative protocols for WPCNs with different setups. In all aforementioned papers, only AP is considered as the energy source of the whole network. In \cite{6697937}, the authors proposed the idea of deploying dedicated power nodes, named power beacons (PBs), to enable WET in the DL. By resorting to the stochastic geometry theory, the densities and transmit power of PBs are investigated under data links' outage constraint. With this PB-based WET, we thus could consider a new network setup, named ``PB-assisted WPCN (PB-WPCN)", in which each user can harvest wireless energy not only from the AP but also from the deployed PB. For this new model, a natural question that arises is ``\textit{how to optimally allocate the resources of PB-WPCNs, including the PBs' energy, and the time for DL WET as well as the UL information~transmission?}". To the best of our knowledge, this is still an open question, which motivates this~paper.

In this paper, we consider a PB-WPCN consisting of one multi-antenna PB and multiple single-antenna AP-source pairs. In view of the state of art and trend of RF energy transfer \cite{lu2014wireless}, the considered network setup is very likely to find its applications in the practical scenario of small cells, such as picocells (range from 10 to 100 meters) and femtocells (WiFi-like range), which has been regarded as one of the key enabling technologies of the upcoming 5G cellular networks \cite{andrews2014will}. It is also worth mentioning that although introducing the PB may result in some extra cost and complexity to the system, this could be beneficial as a whole based on the following considerations: (1) The PB could be dedicated designed for WET only and thus can achieve a higher WET efficiency by exploiting the benefits of energy beamforming enabled by multiple antennas \cite{Enhancingwirelssinfor,6568923}. (2) The deployment of the PB could be more flexible since it has much looser backhaul requirement. (3) By using both the AP and PB to perform WET simultaneously, the transmit diversity could be achieved and thus could make the RF energy harvesting at the user side more robust. We consider that the APs and the PB are connected to constant power supplies. Each AP aims to collect the information from its associated source. It is assumed that each source has no embedded energy supply but has the ability to harvest and store the energy from RF signals broadcast by its AP and/or the PB. The PB is installed to assist APs during the DL WET phase by providing wireless charging service. Note that the energy allocation of the PB and time allocation of each AP-source pair are tangled together. This makes the distributed and optimal resource allocation for the considered PB-WPCN non-trivial at all. The main contributions of this paper are two-fold.

\textit{Firstly}, we consider a cooperative scenario, where all the AP-source pairs and the PB are cooperative to maximize the network social welfare, which is defined as the weighted sum-throughput of all AP-source pairs. This is done by jointly optimizing the DL WET time of each AP and the energy allocation of the PB. A water-filling based algorithm is subsequently proposed to optimally solve the formulated social welfare maximization problem in a distributed~manner. Note that this cooperative scenario can correspond to the situation that APs and the PB are deployed by the same operator, as the one considered in \cite{6697937}.

\textit{Secondly}, we consider a non-cooperative scenario, where all the APs and the PB are assumed to be rational and self-interested. This scenario could be used to model the case that the APs and the PB are installed by different operators. The PB will ask a monetary payment on its wireless charging service and the APs will measure its benefits obtained from the achievable throughput with its payment to the PB. The conflicting interactions among the APs and the PB can be modeled by game theory. Specifically, we formulate an auction game for this non-cooperative scenario based on the well-known Ausubel auction \cite{ausubel2004efficient}, which induces truthful bidding and achieves the maximum social welfare \cite{krishna2009auction,ausubel2004efficient,cramton1998ascending,zhang2011improve}. In this formulated game, the APs are bidders and the PB is the seller as well as the auctioneer. An auction based distributed algorithm is proposed to analyze the formulated game and its convergence is subsequently proved. Note that auction theory \cite{krishna2009auction} has been well investigated and widely applied for designing the resource allocation in cognitive radio networks \cite{5555888}, D2D communication networks \cite{xu2013efficiency}, cooperative communication networks \cite{huang2008auction} as well as energy harvesting networks \cite{ding2014power,niyato2014competitive}. A cooperative network with multiple source-destination pairs and a relay was considered in \cite{ding2014power}, where an auction based power allocation scheme was proposed to allocate the harvested energy of the relay in a distributed manner. \cite{niyato2014competitive} formulated a non-cooperative game to model the competitive WET bidding of the users in a WPCN with one AP and multiple users, where the AP adopted auction mechanism for DL WET. Both \cite{ding2014power} and \cite{niyato2014competitive} adopted the concept of Nash equilibrium to evaluate the strategic interactions between bidders only. In contrast, in this paper we adopt a different auction mechanism to study the hierarchal interaction between the PB (auctioneer) and the AP-source pairs (bidders).

The rest of this paper is organized as follows. We first elaborate the system model in Section~II. Then, a social welfare maximization problem for the cooperative scenario is formulated and a water-filling based distributed algorithm is proposed in Section III. In section IV, an auction game for the non-cooperative scenario is designed and an auction based distributed algorithm is proposed. Section V presents the numerical results. Finally, Section VI concludes the paper.
\begin{figure}
\centering \scalebox{0.27}{\includegraphics{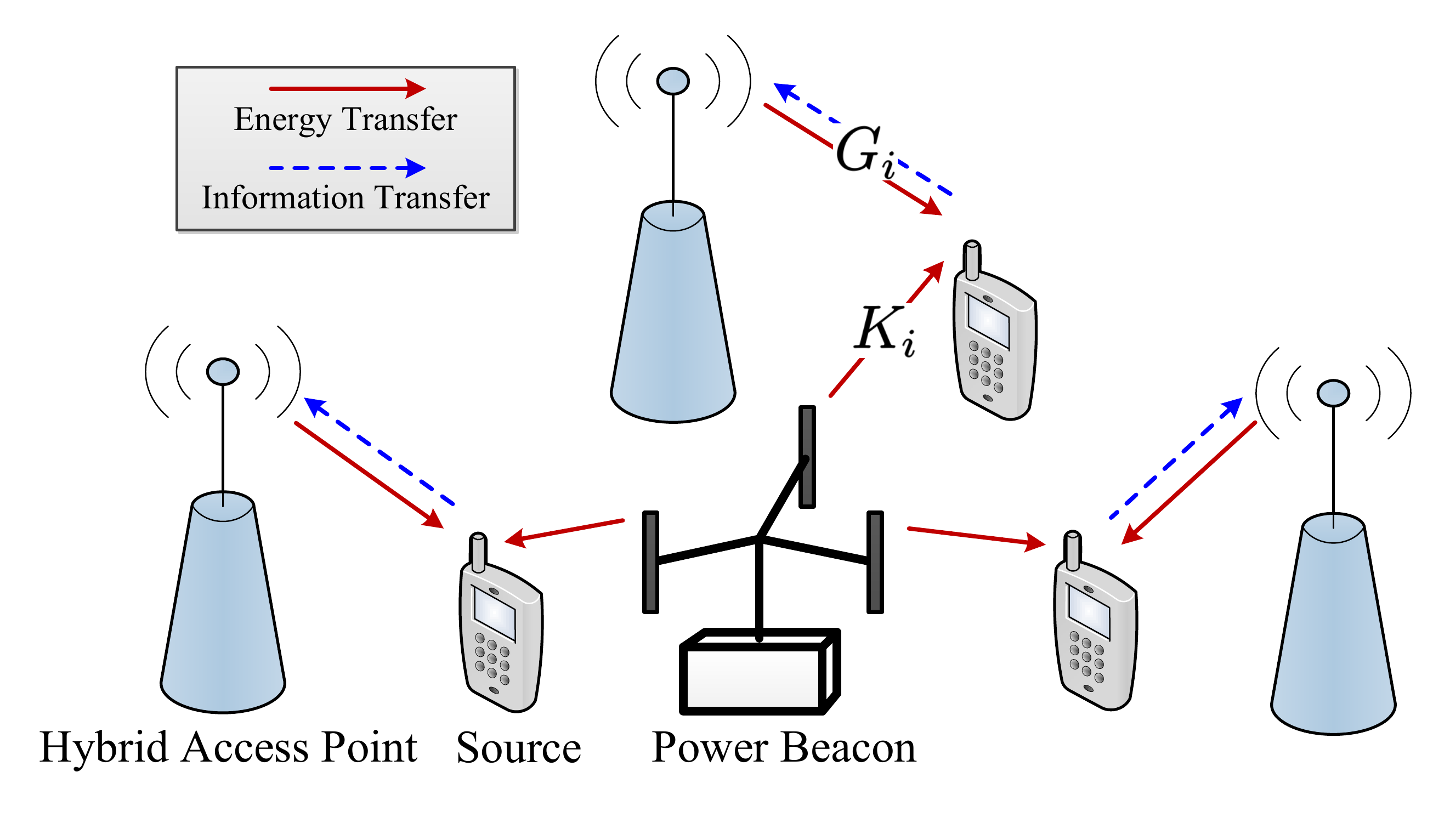}}
\caption{System model for the proposed PB-WPCN.}\label{system}
\end{figure}

\section{System Model}
In this paper, as shown in Fig. \ref{system}, we consider a PB-WPCN with one PB and $N$ APs associated with $N$ (information) sources. Each AP aims to collect the information from its associated source. We denote by $\mathcal{N} =\{1, \cdots, N \}$ the set of AP-source pairs and the $i$th AP-source pair consists of the $i$th AP and the $i$th (associated) source, $\forall i\in\mathcal{N}$. We consider that each source has no fixed energy supplies and thus needs to replenish energy from the signal\footnote{Note that the energy signal could be designed as a zero-mean pseudo-random signal with arbitrary distribution as long as its power spectral density satisfies certain regulations on radio signal radiation for the operating band of interest since it does not carry any intentional information \cite{CollaborativeWireless}.} sent by its AP and/or the PB. The APs and the PB are connected to constant power supplies. We also consider that all APs are connected with the PB via backhaul but they are not connected with each other directly.
We assume that all the AP-source pairs work on orthogonal frequency bands, while the PB can work on all frequency bands. All AP-source pairs work in the half-duplex mode. The APs and sources are each equipped with one single antenna and the PB is equipped with $M  > 1 $ antennas. Full channel state information (CSI) is assumed to be available at the transmitter side.

We exploit the ``harvest-then-transmit" protocol proposed in \cite{6678102}. Specifically, each source first harvests energy from RF signals broadcast by its associated AP and/or the PB in the DL and then uses the harvested energy for information transmission in the UL. By considering that the sources may be low-cost, low-complexity and low-energy devices, we assume that an integrated architecture is adopted at each source, where the energy harvesting component and the information processing component are integrated together by using one rectifier circuit \cite{zhou2013wireless}. In this context, each source can only harvest energy from the RF signals with the frequency inside its working band\footnote{Note that if the sources are advanced enough such that they are equipped with separated front-end hardware for the energy harvesting unit and the information processing unit, it is also interesting to consider the case that the RF energy transfer is performed in a wideband manner and the sources can harvest energy over all frequency bands. However, the design of such a system with wideband energy transfer would be quite different from and much more complicated than the considered one with narrowband energy transfer. Thus, as the initial effort towards the design of the PB-WPCN, we choose to focus on the case with narrowband energy transfer in this paper and we would like to consider the design of another interesting setup with wideband energy transfer as our future work.}, which means that each source can only harvest energy from its own AP and/or the PB.

As illustrated in Fig. 2, we assume that all APs and PB perform WET to sources simultaneously from the beginning of each transmission block $\mathcal{T}$. We denote by $\tau_i$ the fraction of a transmission block $\mathcal{T}$ for DL WET from the $i$th AP to the $i$th source, and we denote by $\tau_i^\prime$ the fraction for DL WET from the PB to the $i$th source. Then the DL WET time for the $i$th source is thus given by $\max\left(\tau_i, \tau_i^\prime\right)  \mathcal{T}$. During the remaining time $\left(1-\max\left(\tau_i, \tau_i^\prime\right)\right)  \mathcal{T}$, the $i$th source uses the harvested energy to transmit its information to its corresponding AP in the UL. For convenience and without loss of generality, we assume that $\mathcal{T} = 1$ in the sequel of~this~paper.

\begin{figure}[!t]
\centering \scalebox{0.4}{\includegraphics{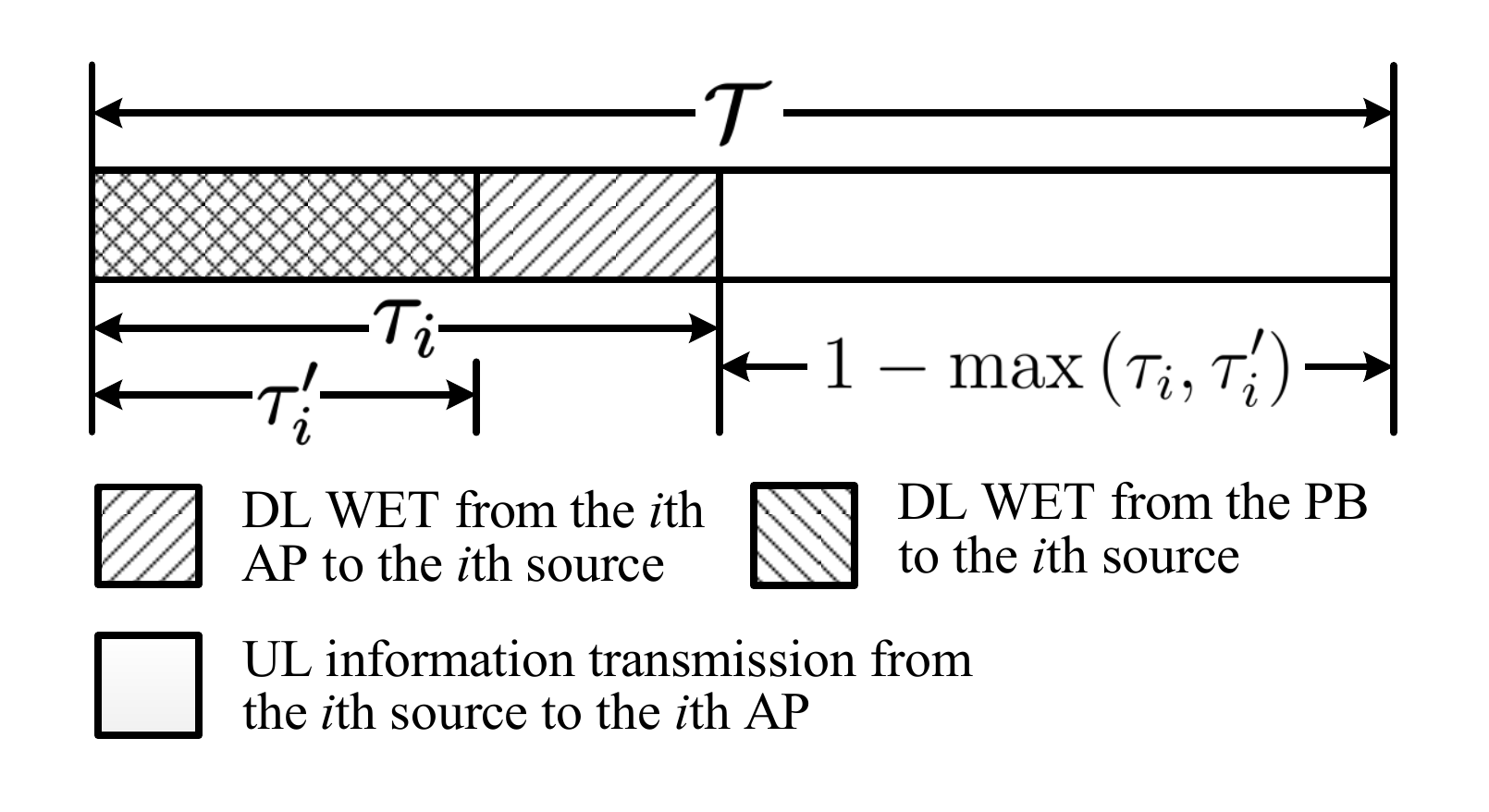}}
\caption{An illustration of the time diagram for $i$th AP-source pair in a transmission block $\mathcal{T}$.}
\end{figure}

We assume that the DL and UL channels are reciprocal. Let scalar $g_i$ denote the complex channel between the $i$th AP and the $i$th source and let $\mathbf{k}_i \in \mathbb{C}^{M\times 1}$ denote the complex channel vector between the PB and the $i$th source, where $\mathbb{C}^{ M \times 1}$ denotes a set of all complex vectors of size $M \times 1$. Besides, we use $\mathbf{w}_i \in \mathbb{C}^{M\times 1}$ to denote the beamforming vector at the PB applied to the energy signal transmitted to the $i$th source with $\|\mathbf{w}_i\|^2 = 1$, where $\|\mathbf{w}_i\|$ is the Euclidean norm of $\mathbf{w}_i$. It is assumed that the $i$th AP transmits with power $p_i$ to the $i$th source and the PB transmits to each source on different frequency bands\footnote{Multi-input single-output and orthogonal frequency division multiplexing (MISO-OFDM) can be adopted at the PB to power the sources on different frequency bands via multiple antennas.} with the same power $p_b$. Furthermore, the PB is constrained by a total energy during each transmission block, denoted by $E_b^{tot}$, which leads to a total energy constraint\footnote{
Generally, an energy consumption constraint should also be imposed to each AP as that to the PB. But, the energy consumption requirement at the APs could be easily satisfied because we consider that the APs are connected to constant power supplies and only serve their own source with fixed transmit power. In contrast, the PB needs to transmit wireless energy to multiple sources and its energy consumption constraint could be frequently violated when the number of sources becomes large. Moreover, it is worth mentioning that the proposed distributed algorithm elaborated in Section III can readily be extended to solve the problem with additional energy constraints for the APs. For the purpose of exposition, in this paper we choose to ignore the energy constraint at each AP.
} $\sum_{i\in\mathcal{N}} \tau_i^\prime p_b \leq E_b^{tot}$.

Then, during the DL WET phase, the received signal at the $i$th source, denoted by $y_i$, can be expressed as follows when both the $i$th AP and the PB perform WET to the $i$th source,
\begin{equation}\label{Eq.0}
\begin{split}
y_i = \sqrt{p_i} g_i x_{i,AP} + \sqrt{p_b}\mathbf{w}_i^\mathrm{H}\mathbf{k}_i x_{i,PB} + n_i,
\end{split}
\end{equation}
where $x_{i,AP}$ and $x_{i,PB}$ are the transmitted signals from the $i$th AP and the PB to the $i$th source, respectively, with $\mathbb{E}\left[\left|x_{i,AP}\right|^2\right] = 1$ and $\mathbb{E}\left[\left|x_{i,PB}\right|^2\right] = 1$. $\mathbb{E}\left[\cdot\right]$ denotes the expectation. $(\cdot)^\textrm{H}$ is the Hermitian transpose. $n_i$ is the additive white Gaussian noise (AWGN) with a zero mean and variance $\sigma^2$. Note that the first or the second term on the right-hand side of (\ref{Eq.0}) should be removed when only the PB or the $i$th AP transfers energy to the $i$th source.

Since the multi-antenna PB transmits energy signal to a single-antenna source and full CSI is available at the PB, the optimal energy beamforming vector at the PB should be maximum ratio transmission (MRT) \cite{6568923,Huang_TVT_2015_On}. We thus have $\mathbf{w}_i = \frac{\mathbf{k}_i}{\|\mathbf{k}_i\|}$. Therefore, at the end of WET phase, the amount of energy harvested by the $i$th source, denoted by $E^s_i$, can be written by
\begin{equation}\label{Eq.1}
\begin{split}
E_{i}^s = \eta  \left( \tau_i p_i G_i + \tau_i^\prime p_b K_i\right),
\end{split}
\end{equation}
where $0 <\eta < 1$ is the energy conversion efficiency, $G_i \triangleq |g_i|^2$ is the channel power gain between the $i$th AP and the $i$th source, $K_i \triangleq \|\mathbf{k}_i\|^2$ is the equivalent channel power gain between the PB and the $i$th source, and the harvested energy from the noise is ignored since it is negligible in practice \cite{zhou2013wireless}. 
We also define that $\boldsymbol{G} \triangleq \left[G_1, \cdots, G_N\right]^\mathrm{H}$ and $\boldsymbol{K} \triangleq \left[K_1, \cdots, K_N\right]^\mathrm{H}$, respectively. Note that the derivation from (\ref{Eq.0}) to (\ref{Eq.1}) is carried out based on the assumption that the signals transmitted by the AP and PB are mutually independent as in \cite{ding2014power,timotheou2014beamforming,chen2014distributed}.

After the source replenishes its energy in the DL, it then transmits its information to the AP in the UL. It is assumed that the harvested energy is exhausted by the source for information transmission. The effect of energy storage and energy consumption of the circuit is disregarded for clarity as in \cite{6678102,30}. The transmit power of the $i$th source, denoted by $q_i$, is thus equal~to
\begin{equation}
\begin{split}
q_i = \frac{E^s_i}{1-\max\{\tau_i,\tau_i^\prime\}} = \frac{\eta  \left( \tau_i p_i G_i+ \tau_i^\prime p_b K_i \right)}{1-\max\left(\tau_i, \tau_i^\prime\right)}.
\end{split}
\end{equation}
Then, the achievable throughput at the $i$th AP can be written~as
\begin{equation}\label{Eq.rate1}
\begin{split}
R_i\left(\tau_i, \tau_i^\prime\right) =& \left(1-\max\left(\tau_i, \tau_i^\prime\right)\right) W \log_2 \left(1 +  \frac{G_i q_i}{\sigma^2}   \right)\\
    = & \left(1-\max\left(\tau_i, \tau_i^\prime\right)\right) W  \\
    &\times \log_2 \left(  1+ \frac{G_i \eta  \left( \tau_i p_i G_i + \tau_i^\prime p_b K_i \right)}{\left(1-\max\left(\tau_i, \tau_i^\prime\right)\right)\sigma^2}    \right),
\end{split}
\end{equation}
where $W$ is the bandwidth and $\sigma^2$ is the noise power, which are assumed to be the same for all AP-source pairs, without loss of~generality. It can be observed from (\ref{Eq.rate1}) that the achievable throughput for the $i$th source to its AP can be increased with the assistance of the PB.

In this paper, we consider both cooperative and non-cooperative scenarios based on whether the PB is cooperative with the APs or not. In the cooperative scenario, the PB and the APs cooperate to maximize the network social welfare, defined as the weighted sum-throughput of all AP-source pairs. In contrast, the PB and the APs are considered to be self-interested in the non-cooperative scenario. More specifically, the PB will ask a monetary compensation for its wireless charging service and each AP-source pair will value its benefits with its payment to the PBs. We will design the resource allocation schemes for these two scenarios in the following sections,~respectively.

\section{Cooperative Scenario}

In this section, we first formulate a social welfare maximization problem for the cooperative scenario and then solve it with a water-filling based algorithm in a distributed manner.

\subsection{Problem Formulation}
We consider a weight $\lambda_i > 0$ for the $i$th AP, which represents a gain per unit throughput from the $i$th source to its AP. Then, the social welfare maximization problem can be formulated~as
\begin{equation}\label{Eq.SW}
\begin{split}
& ~~~~\max_{\left\{\tau_i\right\},\left\{\tau_i^\prime\right\}} ~\sum_{i\in\mathcal{N}} \lambda_i R_i\left(\tau_i, \tau_i^\prime \right),\\
\textrm{s.t.}~&\tau_i, \tau_i^\prime \geq 0,~0 < \max\left(\tau_i, \tau_i^\prime\right)<1,~ \forall i\in\mathcal{N},\\
& ~~~~~~~~~~\sum_{i\in\mathcal{N}}\tau_i^\prime p_b \leq E_b^{tot},
\end{split}
\end{equation}
where the constraint $0 < \max\left(\tau_i, \tau_i^\prime\right)<1$ guarantees that the weighted sum-throughput is larger than zero.
By observing the objective function of problem (\ref{Eq.SW}), we obtain the following lemma:
\begin{lemma}\label{Lem1}
The function $R_i\left(\tau_i, \tau_i^\prime \right)$ in (\ref{Eq.rate1}) can be rewritten as
\begin{equation}\label{Eq.rate}
\begin{split}
R_i\left(\tau_i,E_i\right) = \left(1-\tau_i\right) W \log_2 \left(  1+ \frac{G_i \eta  \left( \tau_i p_i G_i + E_i K_i \right)}{\left(1-\tau_i\right)\sigma^2}    \right),
\end{split}
\end{equation}
where $E_i \triangleq \tau_i^\prime p_b$ is the amount of energy that the PB allocates to the $i$th source.
\end{lemma}
\textit{Proof :} We first prove that $\max\{\tau_i, \tau_i^\prime\} = \tau_i$ by contradiction. If $\tau_i < \tau_i^\prime$, then $\max\{\tau_i, \tau_i^\prime\} = \tau_i^\prime$ and (\ref{Eq.rate1}) becomes
 \begin{equation}\label{Eq.6}
\begin{split}
R_i\left(\tau_i,\tau_i^\prime\right)=\left(1-\tau_i^\prime\right) W \log_2 \left(  1+ \frac{G_i \eta  \left( \tau_i p_i G_i + \tau_i^\prime p_b K_i \right)}{\left(1-\tau_i^\prime\right)\sigma^2}    \right).
\end{split}
\end{equation}
We can easily see that the right-hand side of (\ref{Eq.6}) is a monotonically increasing function of $\tau_i$. This means that the achievable throughput $R_i$ can always be enhanced by increasing the value of $\tau_i$ to that of $\tau_i^\prime$. This contradicts the assumption~that $\tau_i < \tau_i^\prime$. We thus have $\tau_i \geq \tau_i^\prime$, i.e., $\max\{\tau_i, \tau_i^\prime\} = \tau_i$. Furthermore, for the ease of presentation, we define that $E_i \triangleq \tau_i^\prime p_b$, which is the amount of energy~that the PB allocates to the $i$th source. Applying these two operations into (\ref{Eq.rate1}), we can rewrite it as (\ref{Eq.rate}), which completes this proof.\hfill$\blacksquare$

Based on Lemma \ref{Lem1}, we can reformulate the problem (\ref{Eq.SW}) as
\begin{equation}\label{Eq.SWO}
\begin{split}
\max_{\boldsymbol{\tau},\boldsymbol{E}} ~&\sum_{i\in\mathcal{N}} \lambda_i R_i\left(\tau_i, E_i \right),\\
\textrm{s.t.}~&\boldsymbol{0} \prec \boldsymbol{\tau} \prec \boldsymbol{1},\\
&\boldsymbol{0}\preceq \boldsymbol{E} \preceq \boldsymbol{\tau} p_b,\\
&\sum_{i\in \mathcal{N}} E_i \leq E_b^{tot},
\end{split}
\end{equation}
where $\boldsymbol{\tau}\triangleq\left[\tau_1, \cdots, \tau_N \right]^T$ and $\boldsymbol{E}\triangleq\left[E_1, \cdots, E_N \right]^T$, the symbols $\prec$ and $\preceq$ represent the element-wise inequality, $\boldsymbol{0}$ or $\boldsymbol{1}$ is a vector of zeros or ones that has the same size as $\boldsymbol{\tau}$ and $\boldsymbol{E}$, and the constraint of $\boldsymbol{E}$ is derived based on $E_i = \tau_i^\prime p_b$ and $0\leq \tau_i^\prime \leq \tau_i$. Moreover, we denote by $\left(\boldsymbol{\tau}^*, \boldsymbol{E}^*\right)$ the optimal solution to the problem (\ref{Eq.SWO}).

We can see from (\ref{Eq.SWO}) that the social welfare can only be maximized by jointly designing the DL WET time of each AP and the energy allocation of the PB. It is worth noting that $\boldsymbol{\tau}$ and $\boldsymbol{E}$ are mutually interdependent and the achievable throughput of all APs are~coupled together~due~to~the~total~energy~constraint~$\sum_{i\in \mathcal{N}} E_i \leq E_b^{tot}$.

\subsection{Optimal Solution and Distributed Algorithm}

The convexity of the objective function in problem (\ref{Eq.SWO}) can readily be checked
by introducing a new optimization variable $\theta_{i} = 1- \max\{\tau_i, \tau_i'\}$, $\forall i\in \mathcal{N}$. Substituting $\theta_i$ into (\ref{Eq.rate1}), we can observe that $R_i\left(\tau_i, \tau_i^\prime\right)$ is jointly concave on $\tau_i$, $\tau_i'$ and $\theta_i$, since $\theta_i$ is the perspective variable of the $\log$. Accordingly, the problem (\ref{Eq.SWO}) is convex and could be solved by applying standard convex optimization approaches. However, these approaches are normally done in a centralized manner. In practice, a distributed approach is of more interest because it can significantly reduce the network overhead.
Motivated by this, we propose a distributed method with three steps to resolve it in this section. In step 1, we first find the optimal relationship between $\tau_i$ and $E_i$ by expressing $\tau_i$ as a function of $E_i$. Then, the problem (\ref{Eq.SWO}) can be reformulated as a problem with only $\boldsymbol{E}$ as a variable. In step 2, we investigate the properties of the reformulated problem. Finally, a water-filling based algorithm is proposed to find the optimal solution of the new problem in step~3. We describe the details of these three steps in the following subsections.

\subsubsection{\textbf{Step 1.} Problem Reformulation}
We first derive the expression of $\tau_i$ as a function of $E_i$ for each AP. With a given $\boldsymbol{E}$ that satisfies $\boldsymbol{0}\preceq \boldsymbol{E} \prec \boldsymbol{1} p_b$ and $\sum_{i\in \mathcal{N}} E_i \leq E_b^{tot}$, the problem (\ref{Eq.SWO}) is decoupled and we can have the following optimization problem for each AP regarding $\tau_i$,
\begin{equation}\label{Pro.tau}
\begin{split}
\max_{{\tau_i}} ~\mathcal{S}_i\left( \tau_i\right),~\textrm{s.t.}~0 < {\tau_i} < {1},~\tau_i \geq \frac{E_i}{p_{b}},
\end{split}
\end{equation}
where $\mathcal{S}_i\left( \tau_i \right)$ can be expressed by
\begin{equation}\label{Eq.tauE}
\begin{split}
\mathcal{S}_i\left( \tau_i \right) &= \lambda_i R_i\left(\tau_i, E_i\right)\\
&= \lambda_i  W \left(1-\tau_i\right) \log_2 \left(  1+ \frac{G_i \eta \left(\tau_i p_i G_i + E_i K_i \right)}{\left(1-\tau_i\right)\sigma^2}    \right).
\end{split}
\end{equation}
We denote by $\tau_i(E_i)$ the~optimal solution of (\ref{Pro.tau}), which is given in the following~proposition.
\begin{proposition}\label{Pro1}
Given $E_i \in [0, p_b)$, the optimal solution $\tau_i\left(E_i\right)$ to the problem (\ref{Pro.tau}) can be expressed by
\begin{equation}\label{Eq.pro1}
\tau_i\left(E_i\right)=
\begin{cases}
\frac{ \left(z_i^\dag - 1\right) \sigma^2 -G_i\eta E_{i} K_i }{{\left(z_i^\dag - 1\right) \sigma^2 +  G_i^2 \eta p_i}}, &\mbox{if ~$ 0 \leq E_{i} \leq E_{i}^{lim}$},\\
\frac{E_{i}}{p_b}, &\mbox{if ~$ E_i^{lim} <E_{i} < p_b$},
\end{cases}
\end{equation}
where
\begin{equation}\label{E_lim}
\begin{split}
E_i^{lim} \triangleq \frac{p_b \left(z_i^\dag -1\right) \sigma^2 }{\left(z_i^\dag -1\right)\sigma^2 + G_i\eta \left(p_i G_i + p_b K_i\right)},
\end{split}
\end{equation}
and ${z_i}^\dag > 1$ can be expressed by
\begin{equation}\label{z_dag}
\begin{split}
z_i^\dag = \exp\left(\mathcal{W}\left( \frac{G_i^2 \eta p_i  - \sigma^2}{\sigma^2 \exp(1)}\right) +1\right),
\end{split}
\end{equation}
in which $\mathcal{W}\left(x\right)$ is the Lambert $\mathcal{W}$ function that is the solution to the equality $x = \mathcal{W}\exp(\mathcal{W})$.
\end{proposition}
\textit{Proof :} See Appendix A. \hfill$\blacksquare$

Then, by replacing $\tau_i$ in $\mathcal{S}_i\left(\tau_i\right)$ with $\tau_i\left(E_i\right)$ given in Proposition \ref{Pro1}, we have a function $\mathcal{S}_i\left(E_i\right)$ with only $E_i$ as the variable given by the following Lemma.
\begin{lemma}\label{Lem2}
$\mathcal{S}_i\left( E_i \right)$ can be expressed by
\begin{equation}\label{Eq.S(E)}
\begin{split}
&\mathcal{S}_i\left( E_i \right)=\\
&\begin{cases}
    \frac{\lambda_i W G_i \eta \left( p_i G_i + E_i K_i\right)}{z_i^\dag \sigma^2 \ln 2}, &\mbox{if ~$ 0 \leq E_{i} \leq E_{i}^{lim}$},\\
    \lambda_i  W \left(1-\frac{E_i}{p_b}\right) \log_2 \left( 1+  \frac{ X_i E_i   }{ p_b- E_i}   \right),  &\mbox{if ~$ E_i^{lim} < E_{i} < p_b$},
\end{cases}
\end{split}
\end{equation}
where $X_i \triangleq \frac{G_i\eta\left(p_i G_i +p_b K_i\right)}{\sigma^2}$.
\end{lemma}
\textit{Proof :} See Appendix B. \hfill$\blacksquare$

Hence, we now can reformulate the problem (\ref{Eq.SWO}) with only $\boldsymbol{E}$ as the variable as follows
\begin{equation}\label{Eq.SWO1}
\begin{split}
\max_{\boldsymbol{E}}~\sum_{i \in \mathcal{N}}\mathcal{S}_i\left( E_i \right),~\textrm{s.t.}~\boldsymbol{0} \preceq \boldsymbol{E} \prec \boldsymbol{1}p_b ,~\sum_{i\in \mathcal{N}} E_i \leq E_b^{tot}.
\end{split}
\end{equation}
Note that we can solve the problem (\ref{Eq.SWO1}) to obtain $\boldsymbol{E}^\star$, and then we can calculate $\boldsymbol{\tau}^\star$ based on (\ref{Eq.pro1}), thereby solving the original problem (\ref{Eq.SWO}).

\subsubsection{{\textbf{Step 2.} Property Characterization}}

To solve the problem (\ref{Eq.SWO1}), we first characterize the properties of the objective function $\sum_{i\in\mathcal{N}}\mathcal{S}_i\left(E_i\right)$. Since $\sum_{i\in\mathcal{N}}\mathcal{S}_i\left(E_i\right)$ is a positive summation of $N$
independent functions with the same structure, we then first investigate the property of any function $\mathcal{S}_i\left(E_i\right)$, which is summarized in the following proposition:

\begin{proposition}\label{Pro2}
When $0 \leq E_i < p_b$, $\mathcal{S}_i\left( E_i \right)$ is differentiable and the gradient of $\mathcal{S}_i\left(E_i\right)$, denoted by $\nabla \mathcal{S}_i\left(E_i\right)$, is continuous. $\nabla \mathcal{S}_i\left(E_i\right)$ is derived~as
\begin{equation}\label{grad}
\nabla \mathcal{S}_i\left(E_i\right)=
\begin{cases}
\alpha_i,  &\mbox{if ~$ 0\leq E_{i} \leq E_{i}^{lim}$},\\
 \beta_i\left(E_i\right), &\mbox{if ~$ E_i^{lim} < E_{i} < p_b$},
\end{cases}
\end{equation}
where $\alpha_i$ is a constant given by 
\begin{equation}\label{Eq.alpha}
\begin{split}
\alpha_i = \frac{ \lambda_i W G_i\eta K_i}{z_i^\dag \sigma^2 \ln 2},
\end{split}
\end{equation}
and $\beta_i\left(E_i\right)$ is strictly decreasing, which is expressed as 
\begin{equation}\label{Eq.beta}
\begin{split}
\beta_i\left(E_i\right) = &-\frac{\lambda_i W}{p_b}\log_2\left(  1+  \frac{X_i E_i   }{ p_b- E_i}   \right) \\
&+ \frac{\lambda_i W X_i }{\left(p_b - E_i + X_i E_i\right)\ln2}.
\end{split}
\end{equation}
\end{proposition}
\textit{Proof :}  See Appendix B. \hfill$\blacksquare$

Therefore, based on Proposition 2, it is easy to show that $\mathcal{S}_i\left(E_i\right)$ is a concave function of $E_i$ when $0 \leq E_i < p_b$. Then, by letting the first-order derivative of $\mathcal{S}_i\left(E_i\right)$ be equal to zero, i.e., $\beta_i\left(E_i\right) = 0$, we obtain the stationary point that maximizes $\mathcal{S}_i\left(E_i\right)$, denoted by $E_i^o$, given by
\begin{equation}\label{E_o}
\begin{split}
E_i^o =\frac{ p_b \left(z_i^\ddag -1\right)\sigma^2}{\left(z_i^\ddag -1\right)\sigma^2 + G_i\eta\left(p_i G_i +p_b K_i\right)},
\end{split}
\end{equation}
where $z_i^\ddag > 1$ can be expressed as
\begin{equation}
\begin{split}
z_i^\ddag = \exp\left(\mathcal{W}\left( \frac{G_i\eta\left(p_i G_i +p_b K_i\right) - \sigma^2}{ \sigma^2\exp(1)}\right) +1\right).
\end{split}
\end{equation}
Note that $E_i^{o}$ has the same structure as $E_i^{lim}$ given in (\ref{E_lim}). The only differences between them are the parameters $z_i^{\ddag}$ and $z_i^{\dag}$. Furthermore, we can easily observe that $z_i^{\ddag} > z_i^{\dag} >1$ based on the property of the equation (\ref{A.2}) in Appendix A. Let us divide both the top and the bottom of the right-hand side in (\ref{E_o}) by $\left(z_i^\ddag -1\right)\sigma^2$ ( in (\ref{E_lim}) by $\left(z_i^\dag -1\right)\sigma^2$). We thus can obtain that $E_i^{lim} <E_i^o < p_b$. Now, we have an important property of $\mathcal{S}_i\left(E_i\right)$: it is increasing when $0 \leq E_i \leq E_i^o$ and decreasing when $E_i^o < E_i  < p_b$. The PB will at most allocate the amount of energy that is equal to $E_i^o$ to the $i$th AP to achieve the maximization of the social welfare. We thus only need to focus on the interval $0  \leq {E_i} \leq {E_i^o}$ for each AP. We now can update the property of the function $\mathcal{S}_i\left(E_i\right)$ over the interval $0  \leq {E_i} \leq {E_i^o}$ in the following~proposition:

\begin{proposition}\label{Pro3}
For $0  \leq {E_i} \leq {E_i^o}$, $\mathcal{S}_i\left(E_i\right)$ is differentiable, increasing and concave, and $\nabla \mathcal{S}_i\left(E_i\right)$ is continuous.~$\nabla \mathcal{S}_i\left(E_i\right)$ is constant when $0 \leq E_i \leq E_i^{lim}$ while strictly decreasing when $ E_i^{lim}< E_i < E_i^o$ and equal to zero when $E_i = E_i^o$.
\end{proposition}

\begin{figure*}[t]
\centering
 \subfigure[$\tau_i\left(E_i\right)$ versus $E_i$.]
  {\scalebox{0.37}{\includegraphics {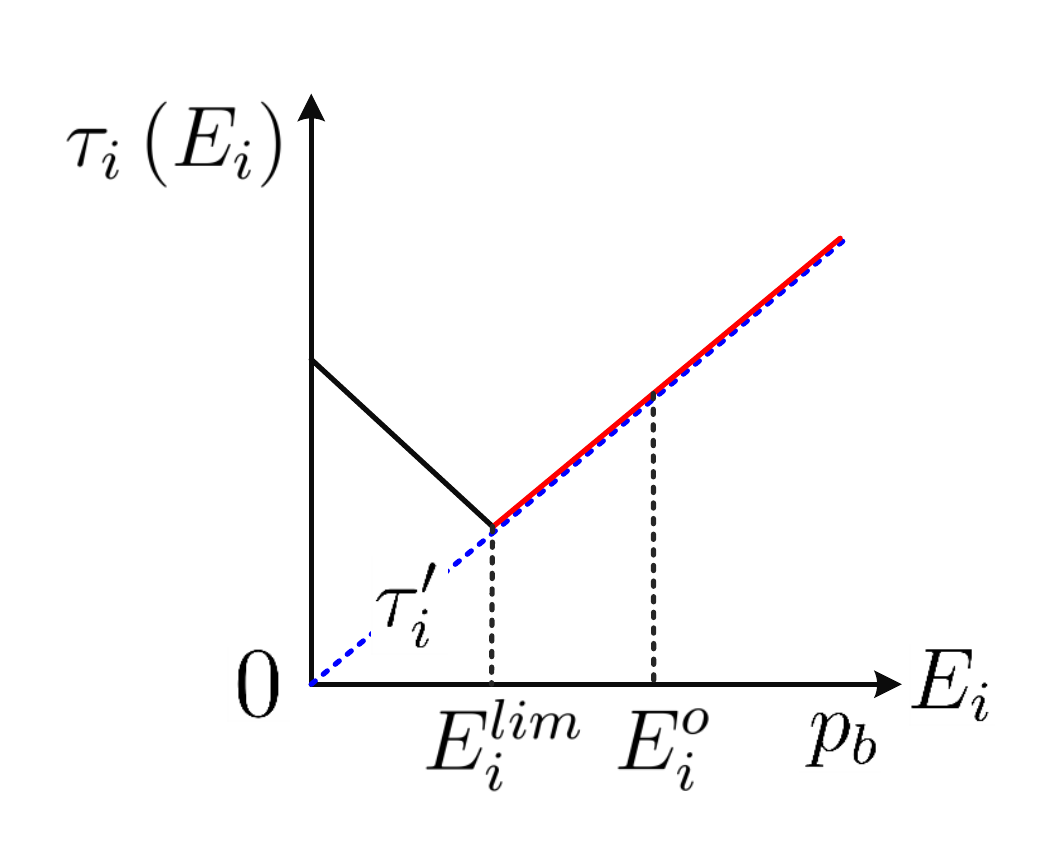}
  \label{tau_i}}}
\hfil
 \subfigure[$\mathcal{S}_i\left(E_i\right)$ versus $E_i$. ]
  {\scalebox{0.37}{\includegraphics {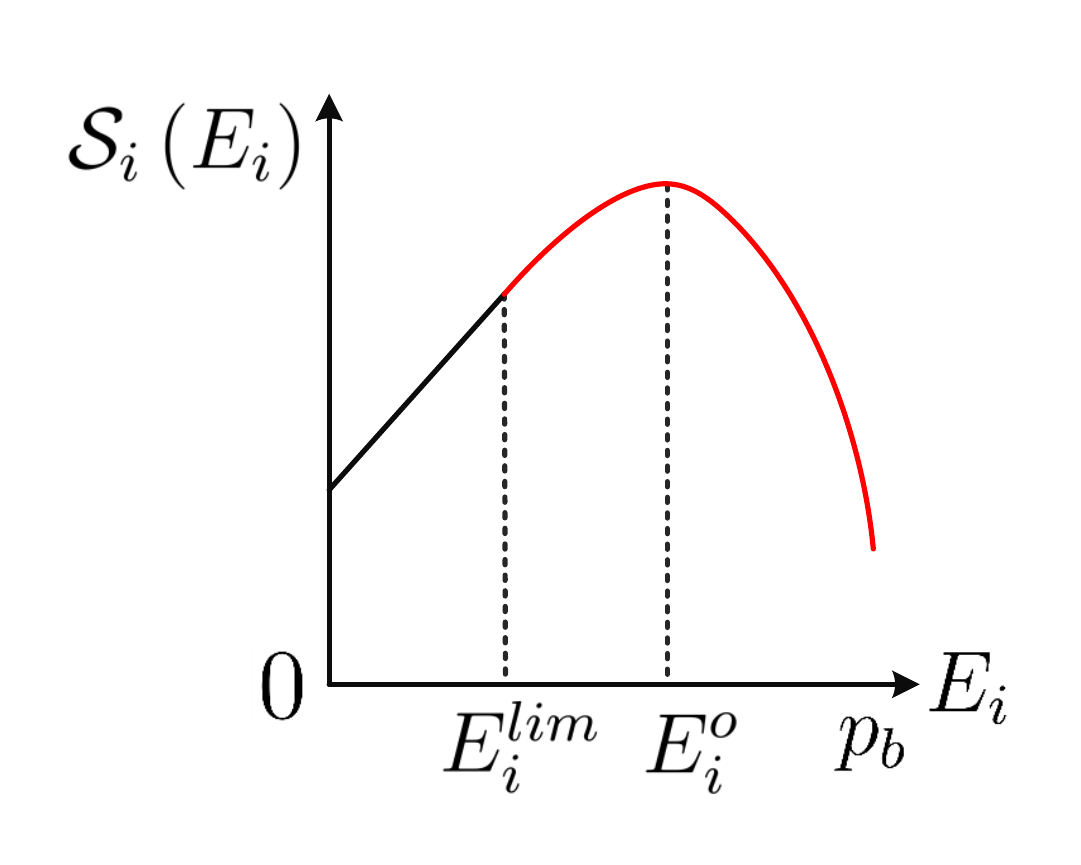}
\label{S_i}}}
\hfil
 \subfigure[$\nabla \mathcal{S}_i\left(E_i\right)$ versus $E_i$.]
  {\scalebox{0.37}{\includegraphics {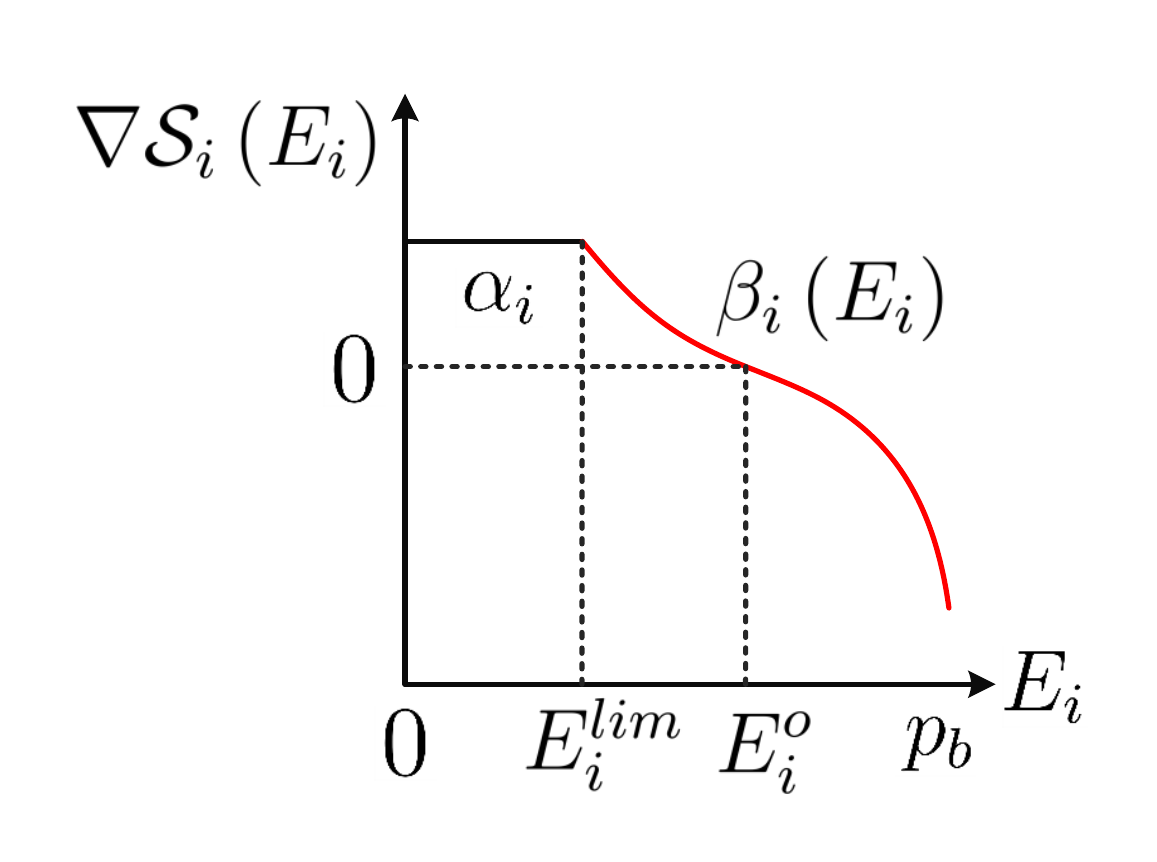}
\label{gradi}}}
\caption{Graphical interpretations for the functions (\ref{Eq.pro1}), (\ref{Eq.S(E)}) and (\ref{grad}) in the proposed cooperative scenario.}
\label{function}
\end{figure*}

For the ease of understanding, functions $\tau_i\left(E_i\right)$, $\mathcal{S}_i\left(E_i\right)$ and $\nabla \mathcal{S}_i\left(E_i\right)$, given in (\ref{Eq.pro1}), (\ref{Eq.S(E)}) and (\ref{grad}), respectively, are graphically interpreted in Fig.~\ref{function}. We use the black and red curves to characterize the functions in the intervals $0 \leq E_i \leq E_i^{lim}$ and $E_i^{lim}<E_i <p_b $, respectively.  We can see from Fig. \ref{tau_i} that $\tau_i\left(E_i\right)$ linearly decreases in the interval $0 \leq E_i \leq E_i^{lim}$, but linearly increases when $E_i^{lim}<E_i <p_b$. The harvesting time from the PB $\tau_i^\prime = \frac{E_i}{p_b}$ is also shown in Fig. \ref{tau_i} using the blue dashed line, which is always linearly increasing in the whole interval and coincides with the red curve in the interval $E_i^{lim}<E_i <p_b$. It can be observed in Fig.~\ref{S_i} that $\mathcal{S}_i\left(E_i\right)$ is linearly increasing in the interval $0 \leq E_i \leq E_i^{lim}$, which corresponds to the observation in Fig.~\ref{gradi} that $\nabla \mathcal{S}_i\left(E_i\right)$ is constant for $0 \leq E_i \leq E_i^{lim}$. With the increasing of $E_i$, $\mathcal{S}_i\left(E_i\right)$ is maximized at the point $E_i^o$, where $\nabla \mathcal{S}_i\left(E_i\right)$ equals zero. Finally, when $E_i^o < E_i < p_b$, $\mathcal{S}_i\left(E_i\right)$ keeps decreasing and $\nabla \mathcal{S}_i\left(E_i\right)$ is shown to be always negative.

\subsubsection{\textbf{Step 3.} A Distributed Algorithm}
Based on the analysis in step 2, the optimization problem (\ref{Eq.SWO1}) can be written as
\begin{equation}\label{Eq.SWO2}
\begin{split}
\max_{\boldsymbol{E}} ~&\sum_{i \in \mathcal{N}}\mathcal{S}_i\left( E_i \right),\\
\textrm{s.t.}~&\boldsymbol{0} \preceq\boldsymbol{E} \preceq \boldsymbol{E^o},\\
&\sum_{i\in \mathcal{N}} E_i \leq E_b^{tot},
\end{split}
\end{equation}
where $\boldsymbol{E^o} \triangleq \left[E_1^o \cdots, E_N^o\right]^T$. Based on the property of $\mathcal{S}_i\left( E_i \right)$ summarized in Proposition~\ref{Pro3}, we then can solve the problem (\ref{Eq.SWO2}) using a water-filling based approach \cite{palomar2005practical}. It should be noted that when $E_b^{tot} \geq \sum_{i\in\mathcal{N}} E_i^o$, the optimization problem (\ref{Eq.SWO2}) has a trivial solution that $E_i^\star = E_i^o,~\forall i\in\mathcal{N}$. We thus only consider the case that $E_b^{tot} < \sum_{i\in\mathcal{N}} E_i^o$, namely, the PB's energy will be exhausted to enhance the social welfare. The optimal solution ${\boldsymbol{E}^\star}$ to the problem (\ref{Eq.SWO2}) can be presented by the following proposition.
\begin{proposition}\label{Pro4}
The optimal solution to the problem (\ref{Eq.SWO2}) is given by ${\boldsymbol{E}^\star} = \left[{E}_1^\star, \cdots, {E}_N^\star\right]^T$~with
\begin{equation}\label{Eq.pro4}
E_i^\star=
\begin{cases}
   \gamma_i\left(\nu\right),  &\mbox{if ~$ 0 \leq \nu < \alpha_i$},\\
   E_b^{tot} - \sum_{j \in \mathcal{N}\backslash \{i\}}E_j^\star, &\mbox{if ~$ \nu = \alpha_i$}\\
    0, &\mbox{if ~$ \nu > \alpha_i$},
\end{cases}
\end{equation}
where $\nu \geq 0$ is a constant chosen to meet the total energy constraint $\sum_{i\in \mathcal{N}} E_i^\star = E_b^{tot}$ and $\gamma_i\left(\nu\right)$ is a strictly decreasing function on $0 \leq \nu < \alpha_i$ given by
\begin{equation}\label{gamma}
\begin{split}
\gamma_i\left(\nu\right) = \frac{p_b \left(z_i^\S -1\right) \sigma^2 }{\left(z_i^\S - 1\right)\sigma^2 + G_i\eta\left(p_i G_i +p_b K_i\right)},
\end{split}
\end{equation}
where $z_i^\S > 1$ is the unique solution of the equation
\begin{equation}\label{gamma2}
\begin{split}
z_i \ln\left(z_i\right) + \left(\frac{\nu p_b\ln2}{\lambda_i W} -1\right)z_i + 1 = \frac{G_i\eta\left(p_i G_i +p_b K_i\right)}{\sigma^2}.
\end{split}
\end{equation}
\end{proposition}
\textit{Proof :} See Appendix C. \hfill$\blacksquare$

According to Proposition \ref{Pro3} and Proposition \ref{Pro4}, there exists a unique $\nu$ to meet the total energy constraint. However, from Proposition \ref{Pro4}, if $\nu = \alpha_i$, then $E_i^\star$ has a unique solution only under the assumption that each AP corresponds to a different $\alpha_i$, $\forall i\in \mathcal{N}$. This is because $E_j^\star$, $\forall j\in\mathcal{N}\backslash\{i\}$ is unique when $\nu \neq \alpha_j$.
If there is at least another AP, say the $k$th AP, $k \in \mathcal{N}\backslash\{i\}$, and $\nu = \alpha_k = \alpha_i$, then the optimality can be achieved with multiple solutions. Based on the proof of Proposition~\ref{Pro4}, the solutions of $E_i^\star$ and ${E}_k^\star$ can be any combination, which is subject to that $0 \leq E_i^\star \leq E_i^{lim}$, $0 \leq {E}_k^\star \leq E_k^{lim}$ and $E_i^\star + {E}_k^\star = E_b^{tot} - \sum_{j \in \mathcal{N}\backslash \{i,k\}} E_j^\star$.

Next, we propose a water-filling based distributed algorithm to obtain ${\boldsymbol{E}^\star}$ by finding the constant $\nu$. Note that we only consider the case that each AP corresponds to a different $\alpha_i$, which can be readily modified to adapt to the case that multiple APs have the same $\alpha_i$. For convenience and without loss of generality, we assume that $\alpha_1 > \alpha_2  > \cdots > \alpha_{N}$. As $\nu$ is unique and $\nu \geq 0$, a possible value of $\nu$ could be $\alpha_i$, $\forall i\in\mathcal{N}$, or between an interval $\left(\alpha_{i +1}, \alpha_i\right)$, where~$\alpha_{N+1} \triangleq 0$. Note that when the optimal $\nu$ falls into an interval between any two adjacent $\alpha_i$'s, we need to apply the iterative water-filling algorithm  \cite{yu2007multiuser} to achieve it. Specifically, after the PB releases the current energy price in each round, all the APs update their bids in parallel and feedback them to the PB within the current round.

We now elaborate the steps in Algorithm \ref{Algorithm1}. Each AP first provides its $\alpha_i$ to the PB, then the PB constructs a descending sequence $\{\alpha_i\}_{i\in\mathcal{N}}$, sets $i = 1$ and announces $\nu^{(i)}$, which is equal to each $\alpha_i$, sequentially. Then, with $\nu^{(i)}$, the $j$th AP, $\forall j \in \mathcal{N}\backslash\{i\}$, computes the response, here denoted by $E_j^{(i)}$, according to (\ref{Eq.pro4}), while\footnote{According to Proposition \ref{Pro4}, given $\nu = \alpha_i$, the response energy for the $i$th AP is that $E_i^\star =  E_b^{tot} - \sum_{j \in \mathcal{N}\backslash \{i\}}E_j^\star$. However, this happens only when $\nu$ is found to meet the total energy constraint. As proved in Appendix C, given $\nu = \alpha_i$, under the assumption that the $i$th AP does not know the total energy constraint, the possible value of $E_i^\star$ exists in an interval $\left[0, E_i^{lim}\right]$. Therefore, in this case, the $i$th AP is set to respond that $E_i^{(i)} = E_i^{lim}$ to the PB. Then the PB knows the interval of the $i$th AP, which will help it to make decisions in the following steps.} the $i$th AP responds that $E_i^{(i)} = E_i^{lim}$. Then, by comparing the aggregate of all the responses, computed by $E^{(i)}_{agg} = \sum_{j\in\mathcal{N}\setminus\{i\}}E_j^{(i)} + E_{i}^{(i)}$, with $E_b^{tot}$, the PB can decide the value of $\nu$ as listed in Algorithm~\ref{Algorithm1}. Note that if $\nu$ exists in an interval, then a bisection method \cite{wilkinson1965algebraic} can be employed to find the unique value. The implementation of the bisection method is simple and thus~omitted. It can be observed that Algorithm \ref{Algorithm1} can always find the exact value or the existing interval of $\nu$ with at most $N$ iterations, since $\nu$ is unique and $0\leq \nu \leq \max_{i\in\mathcal{N}} \{\alpha_i\}$.

\begin{algorithm}[t]
  \begin{algorithmic}[1]
    \State Each AP reflects its $\alpha_i$ to the PB.
    \State The PB constructs a descending sequence $\{\alpha_i\}_{i\in\mathcal{N}}$, sets $i = 1$ and repeats:
    \begin{enumerate}
      \item The PB announces $\nu^{(i)} = \alpha_i$ to all APs.
      \item The $j$th AP responds an optimal $E_j^{(i)}$ to the PB according to Proposition \ref{Pro4}, $\forall j \in \mathcal{N}\backslash \{i\}$, while the $i$th AP responds $E_i^{(i)} = E_i^{lim}$.
      \item The PB computes that $E^{(i)}_{agg} = \sum_{j\in\mathcal{N}\backslash \{i\}}E_j^{(i)} + E_i^{(i)}$ and compares~$E^{(i)}_{agg}$ with $E_b^{tot}$:
          \begin{enumerate}
            \item If $E^{(i)}_{agg} < E_b^{tot}$, then set $i = i + 1$ and continue.
            \item Elseif $E^{(i)}_{agg} \geq E_b^{tot}$ and $E^{(i)}_{agg} - E_i^{lim} \leq E_b^{tot}$, then $\nu = \alpha_i$, $E_j^\star = E_{j}^{(i)}$, and $E_i^\star = E_b^{tot} - \sum_{j \in \mathcal{N}\backslash\{i\}} E_{j}^{(i)}$.
            \item Else,
             the unique $\nu \in (\alpha_{i}, \alpha_{i-1})$, which can be readily found via a bisection method.
          \end{enumerate}
    \end{enumerate}
  \end{algorithmic}
  \caption{Water-filling based Distributed Algorithm}
  \label{Algorithm1}
\end{algorithm}

According to Algorithm \ref{Algorithm1}, each AP first needs to measure the (equivalent) channel power gains $G_i$ and $K_i$, and acquire the transmit power $p_b$ from the PB. The values calculated by each AP and forwarded to the PB should be $\alpha_i$ given in (\ref{Eq.alpha}), $E_i^{lim}$ defined in (\ref{E_lim}) and the optimal response obtained by (\ref{Eq.pro4}). With these signals from the APs, the PB announces $\nu_i^{(i)}$ at the end of each iteration and the allocated energy $E_i^\star$ at last to each AP.

Finally, with the value of $\nu$ achieved via Algorithm \ref{Algorithm1} and (\ref{Eq.pro4}), we can obtain $\boldsymbol{E}^\star$. Subsequently, based on (\ref{Eq.pro1}) given in Proposition \ref{Pro1}, we have $\tau_i^\star = \tau_i\left(E_i^\star\right)$ and $\boldsymbol{\tau}^\star  = \left[\tau_1^\star,\cdots, \tau_N^\star\right]^T$. Therefore, we have already found the optimal solution $\left(\boldsymbol{\tau}^\star,\boldsymbol{E}^\star\right)$ of the original problem (\ref{Eq.SWO}).

\section{Non-cooperative Scenario}
This section investigates the non-cooperative scenario, where the multiple APs and the PB are assumed to be rational and self-interested.
We adopt an auction model to characterize the conflicting interactions among the APs and the PB, where the APs are bidders and the PB is the auctioneer. In particular, we employ a well-known Ausubel auction \cite{ausubel2004efficient}, which constructs an ascending-bid version of the Vickrey-Clarke-Groves (VCG) auction and induces truthful bidding\footnote{Truthful bidding means that reporting true optimal demand at every iteration is a mutually best response for each bidder~\cite{krishna2009auction}.} as well as achieves the maximum social welfare, i.e., the global optimum \cite{krishna2009auction,ausubel2004efficient,cramton1998ascending,zhang2011improve}.
In the sequel, we first define the utility functions of both APs and the PB. Then we formulate an auction game, in which an auction based distributed algorithm is proposed. At last, we analyze the formulated game and prove convergence of the proposed algorithm.

\subsection{Utility functions}

We first present the utility functions of APs and the PB, respectively.

\subsubsection{The Utility Function of Each AP}
Given the unit price of the PB's energy, denoted by $\mu$, the payment from the $i$th AP can be written by
\begin{equation}\label{Eq.payment}
\begin{split}
\Gamma_i =  \mu E_i.
\end{split}
\end{equation}
Therefore, with (\ref{Eq.rate}) and (\ref{Eq.payment}), the utility function of the $i$th AP can be defined as
\begin{equation}\label{Eq.Ua}
\begin{split}
\mathcal{U}_i\left(\tau_i, E_i, \mu\right) =& \lambda_i R_i\left(\tau_i, E_i\right) - \Gamma_i \\
=& \lambda_i (1-\tau_i) W \\
&\times \log_2 \left(  1+ \frac{G_i \eta  \left( \tau_i p_i G_i + E_i K_i \right)}{\left(1-\tau_i\right)\sigma^2}    \right) - \mu E_i.
\end{split}
\end{equation}

\subsubsection{The Utility Function of the PB}

Recall that $\boldsymbol{E} = \left[E_1, \cdots, E_N\right]^T$. Then, the utility of the PB can be expressed as
\begin{equation}\label{U.bec}
\begin{split}
\mathcal{U}_b\left(\mu,\boldsymbol{E}\right) = \sum_{i\in \mathcal{N}} \Gamma_i = \mu \sum_{i\in \mathcal{N}}  E_i.
\end{split}
\end{equation}
Note that, there is a reserve price for the PB, denoted by $\mu^{(0)}$. It is assumed that if the outcome price of an auction game is larger than $\mu^{(0)}$, the PB would benefit from the trade. \textit{Otherwise, it would not participate in the trade} \cite{zhang2011improve,5555888}.

\subsection{Auction Game}

Now, we are ready to formulate an auction game, where the APs are bidders and the PB is the auctioneer. Each AP submits bid to compete for the PB's energy, in order to maximize its utility defined in (\ref{Eq.Ua}).
On the other hand, the PB aims to maximize its revenue in (\ref{U.bec}) by increasing its energy's unit price. In particular, the auctioneer first announces an initial price $\mu = \mu^{(0)}$, and the bidders respond to the auctioneer with their optimal demands, i.e., the bids. Then, the auctioneer raises the price $\mu$ until the aggregate demands meet the total energy constraint, and meanwhile, the auctioneer concludes the auction and decides the final allocated energy to each bidder.

In the formulated game, at each round $t \geq 0$, with a given price $\mu^{(t)}$, the $i$th AP has an optimization problem as
\begin{equation}\label{AP_OP}
\begin{split}
\max_{\tau_{i}, E_i}~ \mathcal{U}_i\left(\tau_i, E_i, \mu^{(t)} \right), ~\textrm{s.t.}~ 0 < \tau_i <1,~ 0 \leq E_i \leq  \tau_i p_b.
\end{split}
\end{equation}
We denote by $\left(\tau_i^{(t)}, E_i^{(t)}\right)$ the optimal solution of (\ref{AP_OP}). That~is
\begin{equation}\label{Eq.opts}
\begin{split}
\left(\tau_{i}^{(t)}, E_{i}^{(t)}\right) = \arg \max_{\tau_i, E_i} ~\mathcal{U}_i\left(\tau_i, E_i, \mu^{(t)}\right).
\end{split}
\end{equation}
Thus the bid of the $i$th AP is $E_i^{(t)}$ given $\mu^{(t)}$. We also denote by $\widetilde{E}_i^\star$ and $\widetilde{\Gamma}_i^\star$ the final allocated energy and the final payment of the $i$th AP when the PB concludes the auction. With $\widetilde{E}_i^\star$ and $\widetilde{\Gamma}_i^\star$, the utility of the $i$th AP in (\ref{Eq.Ua}) becomes to
\begin{equation}\label{Eq.Ua2}
\begin{split}
\mathcal{U}_i\left(\tau_i \right)
&= \lambda_i (1-\tau_i) W\\
& \times \log_2 \left(  1+ \frac{G_i \eta  \left( \tau_i p_i G_i + \widetilde{E}_i^\star K_i \right)}{\left(1-\tau_i\right)\sigma^2}    \right) - \widetilde{\Gamma}_i^\star.
\end{split}
\end{equation}
We then have an optimization problem regarding $\tau_i$ as
\begin{equation}\label{Eq.30}
\begin{split}
\max_{\tau_i}~ \mathcal{U}_i\left(\tau_i \right), ~\textrm{s.t.}~0<\tau_i<1,~\tau_i\geq\frac{\widetilde{E}_i^\star}{p_b}.
\end{split}
\end{equation}
We use $\widetilde{\tau}_i^\star$ to denote the optimal solution to problem (\ref{Eq.30}), which is the optimal harvesting time of the $i$th source from its own AP. Analogue to the proof of Proposition \ref{Pro1}, we can readily get $\widetilde{\tau}_i^\star = \tau_i\left(\widetilde{E}_i^\star\right)$, where $\tau_i\left(\cdot\right)$ is given by (\ref{Eq.pro1}).

\begin{algorithm}[t]
  \begin{algorithmic}[1]
    \State Given $E_b^{tot}$, price step $\Delta > 0$, and $t = 0$, the PB initializes the asking price $\mu^{(0)}$.
    \State With $\mu^{(0)}$, each AP computes and submits its optimal bid ${E_i^{(0)}}$ with (\ref{Eq.opts}).
    \State The PB sums up all bids that $\mathcal{E}^{(0)} = \sum_{i\in \mathcal{N}} {E_i^{(0)}}$ and compares $\mathcal{E}^{(0)}$ with $E_b^{tot}$:
    \begin{enumerate}
      \item If $\mathcal{E}^{(0)} \leq E_b^{tot}$, then the PB ends the auction and quits the trade.
      \item Else, then the PB computes the cumulative clinches $\{\mathcal{C}_{i}^{(0)}\}_{i\in \mathcal{N}}$ by (\ref{Eq.clinch}), sets $\mu^{(t+1)} = \mu^{(t)} + \Delta$, $t= t+1$, and repeats:
      \begin{enumerate}
        \item The PB announces $\mu^{(t)}$ to all APs.
        \item Given $\mu^{(t)}$, each AP updates its optimal bid $E_{i}^{(t)}$ obtained by~(\ref{Eq.opts}).
        \item The PB sums up all bids that $\mathcal{E}^{(t)} = \sum_{i\in \mathcal{N}} E_{i}^{(t)}$ and compares $\mathcal{E}^{(t)}$ with~$E_b^{tot}$:
        \begin{enumerate}
          \item If $\mathcal{E}^{(t)} > E_b^{tot}$, then the PB records the cumulative clinches $\{\mathcal{C}_{i}^{(t)}\}_{i\in \mathcal{N}}$ and sequentially sets $\mu^{(t+1)} = \mu^{(t)} + \Delta$, $t= t+1$. The auction continues.
          \item Else, then by setting $T = t$, the PB concludes the auction and computes $\{\mathcal{C}_{i}^{(T)}\}_{i\in\mathcal{N}}$ with (\ref{Eq.Tclinch}). Then, the PB allocates $\widetilde{E}_i^\star = \mathcal{C}_{i}^{(T)}$ to the $i$th~AP.
        \end{enumerate}
      \end{enumerate}
    \end{enumerate}
  \end{algorithmic}
  \caption{Auction Based Distributed Algorithm}
  \label{Algorithm2}
\end{algorithm}

Algorithm \ref{Algorithm2} elaborates the auction steps. Before starting the auction, the PB sets up the iteration index $t = 0$, the constant price step\footnote{Note that the constant step normally introduces errors to the convergence point \cite{low1999optimization}. However, this is not a problem for the considered auction mechanism with the proportional rationing rule (PRR) \cite{saez2007effects} given in (\ref{Eq.Tclinch}), which guarantees that the total energy of the PB can be entirely allocated~at~last.} $\Delta > 0$, and the initial price $\mu^{(0)}$. The price will be announced to all APs, and each AP will submit its optimal bid ${E_i^{(0)}}$ based on the optimal response given by (\ref{Eq.opts}). Then the PB sums up all bids that $\mathcal{E}^{(0)} \triangleq \sum_{i\in \mathcal{N}} E_i^{(0)}$, and compares $\mathcal{E}^{(0)}$ with $E_b^{tot}$. If $\mathcal{E}^{(0)} \leq E_b^{tot}$, then the PB ends the auction and quits the trade. Otherwise, the PB first computes a cumulative clinch \cite{ausubel2004efficient} for each AP, which is the amount of the PB's energy that each AP is guaranteed to be allocated. The cumulative clinch for the $i$th AP at the round $t \geq 0$ is given by
\begin{equation}\label{Eq.clinch}
\begin{split}
\mathcal{C}_{i}^{(t)} = \max \left\{ 0, E_b^{tot} - \sum_{j\in \mathcal{N}\backslash \{i\}} E_{j}^{(t)} \right\}.
\end{split}
\end{equation}
Then, the PB sets $\mu^{(t+1)} = \mu^{(t)} + \Delta$, $t= t+1$ and updates $\mu^{(t)}$ to all APs.

With the updated $\mu^{(t)}$, each AP submits its optimal bid $E_{i}^{(t)}$ based on the optimal response $\left(\tau_{i}^{(t)}, E_{i}^{(t)}\right)$ obtained by (\ref{Eq.opts}). By comparing the aggregate bids $\mathcal{E}^{(t)} \triangleq \sum_{i\in \mathcal{N}} E_{i}^{(t)}$ with $E_b^{tot}$, if $\mathcal{E}^{(t)} > E_b^{tot}$, then the PB records the cumulative clinches $\{\mathcal{C}_{i}^{(t)}\}_{i\in \mathcal{N}}$ and sequentially sets $\mu^{(t+1)} = \mu^{(t)} + \Delta$, $t= t+1$. The auction continues till $\mathcal{E}^{(t)} \leq E_b^{tot}$. By setting $T = t$, the PB concludes the auction and computes the cumulative clinches $\{\mathcal{C}_{i}^{(T)}\}_{i\in \mathcal{N}}$ according to the proportional rationing rule (PRR) \cite{saez2007effects} by
\begin{equation}\label{Eq.Tclinch}
\begin{split}
 \mathcal{C}_{i}^{(T)}= & E_{i}^{(T)}  + \frac{E_{i}^{(T-1)} - E_{i}^{(T)}   }{ \sum_{i\in \mathcal{N}} E_{i}^{(T-1)} -  \sum_{i\in \mathcal{N}} E_{i}^{(T)} }\\
 & \times \left( E_b^{tot} - \sum_{i \in \mathcal{N}} E_{i}^{(T)} \right),
\end{split}
\end{equation}
where $\mathcal{C}_{i}^{(T)}$ is actually the finally allocated energy to the $i$th AP and it is a sum of its last energy bid and a proportion of the remaining energy $\left( E_b^{tot} - \sum_{i \in \mathcal{N}} E_{i}^{(T)} \right)$. Then, the PB allocates $\widetilde{E}_i^\star =  \mathcal{C}_{i}^{(T)}$ to the $i$th AP. With $\widetilde{E}_i^\star$, the $i$th AP can decide its final harvesting time $\widetilde{\tau}_i^\star$. Hence, the utility of the $i$th AP can be expressed by
\begin{equation}\label{Eq.14}
\begin{split}
& {\mathcal{U}_i} \left(\widetilde{\tau}_i^\star, \widetilde{E}_i^\star \right) = \lambda_i \left(1- \widetilde{\tau}_i^\star\right) W \\
&~~~~~~~~~\times \log_2 \left(  1+ \frac{G_i \eta \left( \widetilde{\tau}_i^\star  p_i G_i + \widetilde{E}_i^\star K_i \right)}{\left(1-\widetilde{\tau}_i^\star\right)\sigma^2}    \right) - \widetilde{\Gamma}_i^\star,
\end{split}
\end{equation}
where the payment from the $i$th AP $\widetilde{\Gamma}_i^\star$ is given by
\begin{equation}\label{Eq.15}
\begin{split}
\widetilde{\Gamma}_i^\star = \mu^{(0)}  \mathcal{C}_{i}^{(0)} +  \sum_{t = 1}^T \mu^{(t)} \left(  \mathcal{C}_{i}^{(t)} -  \mathcal{C}_{i}^{(t-1)} \right).
\end{split}
\end{equation}
Notice that the term $\left(  \mathcal{C}_{i}^{(t)} -  \mathcal{C}_{i}^{(t-1)} \right)$ in (\ref{Eq.15}) is actually the amount of energy that the $i$th AP is guaranteed to be allocated with the announced price $\mu^{(t)}$. Their product will be the corresponding payment of the $i$th AP at the current round. By accumulating these payments generated at each round (from $t = 0$ to $t = T$), we can have achieve total payment of the $i$th AP given in (\ref{Eq.15}). The utility of the PB thus is written~by
\begin{equation}
\begin{split}
\mathcal{U}_b =  \sum_{i \in \mathcal{N}} \widetilde{\Gamma}_i^\star = \sum_{i \in \mathcal{N}} \left( \mu^{(0)}  \mathcal{C}_{i}^{(0)} +  \sum_{t = 1}^T \mu^{(t)} \left(  \mathcal{C}_{i}^{(t)} -  \mathcal{C}_{i}^{(t-1)} \right)\right).
\end{split}
\end{equation}

\subsection{Analysis of the Formulated Game}\label{Sec.IV}
In this subsection, we analyze the formulated auction game.

We first derive the optimal solution to problem (\ref{AP_OP}) given in the following proposition.
\begin{proposition}\label{Pro5}
Given a price $\mu^{(t)}$ by the PB, the optimal solution to problem (\ref{AP_OP}) can be expressed~as
\begin{equation}\label{Eq.36}
\left(\tau_i^{(t)}, E_i^{(t)}\right)=
\begin{cases}
   \left(\frac{ \left(z_i^\dag - 1\right) \sigma^2}{\left(z_i^\dag - 1\right) \sigma^2 +  G_i^2 \eta p_i }, 0 \right), &\mbox{if ~$\mu^{(t)} \geq \mu^{lim}_i$},\\
   \left( \frac{\gamma_i\left(\mu^{(t)}\right)}{p_b}, \gamma_i\left(\mu^{(t)}\right)\right), &\mbox{if ~$\mu^{(t)} < \mu^{lim}_i$},
\end{cases}
\end{equation}
where $\mu_i^{lim}$ is defined as the maximum acceptable price for the $i$th AP, which is equal to $\alpha_i$ in (\ref{Eq.alpha}), $z_i^\dag$ and $\gamma_i\left(\cdot\right)$ are expressed in (\ref{z_dag}) and (\ref{gamma}), respectively.
\end{proposition}
\textit{Proof :} See Appendix D. \hfill$\blacksquare$

Next, we prove that the proposed auction based distributed algorithm has the following convergence property.
\begin{proposition}
The proposed auction based distributed algorithm converges within a finite number of iterations.
\end{proposition}
\textit{Proof :} Recall that $\mathcal{E}^{(t)} = \sum_{i\in \mathcal{N}} E_{i}^{(t)}$ and the auction concludes when $\mathcal{E}^{(t)} \leq E_b^{tot}$. According to Proposition \ref{Pro5}, if the asking price $\mu^{(t)} \geq \mu^{lim}_i$, then the $i$th AP will drop from the auction with $E_i^{(t)} = 0$. If the asking price $\mu^{(t)} < \mu^{lim}_i$, then the $i$th AP always submits a non-zero bid $E_i^{(t)} = \gamma_i\left(\mu^{(t)} \right)$. By referring to Proposition \ref{Pro4}, we can show that $\gamma_i\left(\mu^{(t)} \right)$ is a strictly decreasing function when $0\leq \mu^{(t)} < \mu_i^{lim}$ and when $\mu{(t)}$ approaches $\mu_i^{lim}$, we can get
\begin{equation}
\begin{split}
\lim_{\mu\rightarrow \mu_i^{lim}} E_i^{(t)} = E_i^{lim}.
\end{split}
\end{equation}

We denote by $n$ the number of iterations when the proposed auction concludes. Recall that the price step is $\Delta$. Thus $\mu^{(T)} = \Delta \times  n  + \mu^{(0)}$. When $n$ grows, $\mu^{(T)}$ increases. Since $E_i^{(T)}$ is a decreasing function of $\mu^{(T)}$, the summation of bids $\mathcal{E}^{(T)}  = \sum_{i\in \mathcal{N}} E_i^{(T)}$ will decrease as $\mu^{(T)}$ increases. Therefore, there always exists an $\mathcal{E}^{(T)}$ that satisfies $\mathcal{E}^{(T)} \leq E_b^{tot}$ such that the PB concludes the auction. Hence, the number of iterations $n$ is finite, which completes this proof.~\hfill$\blacksquare$

Last, we analyze the required signaling for the computations in the non-cooperative scenario. According to Algorithm \ref{Algorithm2}, each AP first needs to measure the (equivalent) channel power gains $G_i$ and $K_i$, and acquire the transmit power $p_b$ from the PB. The only value calculated by each AP and forwarded to the PB is the optimal bid $E_i^{(t)}$ obtained by (\ref{Eq.36}). With all the bids from the APs, the PB announces the price $\mu^{(t)}$ at the end of each iteration and the allocated energy the allocated energy $\widetilde{E}_i^\star$ at last to each AP.


\begin{table}
    \caption{System Parameters}\label{tab}
    \centering
    \begin{threeparttable}
      \begin{tabular}{|c||c|}
           \hline
           The bandwidth $W$ & $100$KHz\\
           \hline
           The noise power $\sigma^2$& $-80$dBm\\
           \hline
           The path-loss factor $\zeta$& $2$ \\
           \hline
           The transmit power of the PB $p_b$& $2$Watt \\
           \hline
           The transmit power of each AP $p_i$& $1$Watt \\
           \hline
           The gain per unit throughput for each AP $\lambda_i$& $10$/Mbps\\
           \hline
           The distance between the $i$th source and its AP $d_{A_iS_i}$&$10$m \\
           \hline
           The distance between the $i$th source and the PB $d_{PS_i}$&$10$m \\
           \hline
           The energy conversion efficiency $\eta$& $0.5$\\
           \hline
           The number of antennas at the PB & $4$ \\
           \hline
           The reserve price for the PB $\mu^{(0)}$& $0.001$\\
           \hline
           The price step $\Delta$& $0.01$\\
           \hline
      \end{tabular}
    \end{threeparttable}
\end{table}

\begin{figure}[t]
\centering
 \subfigure[Convergence of the water-filling based distributed algorithm.]
  {\scalebox{0.35}{\includegraphics {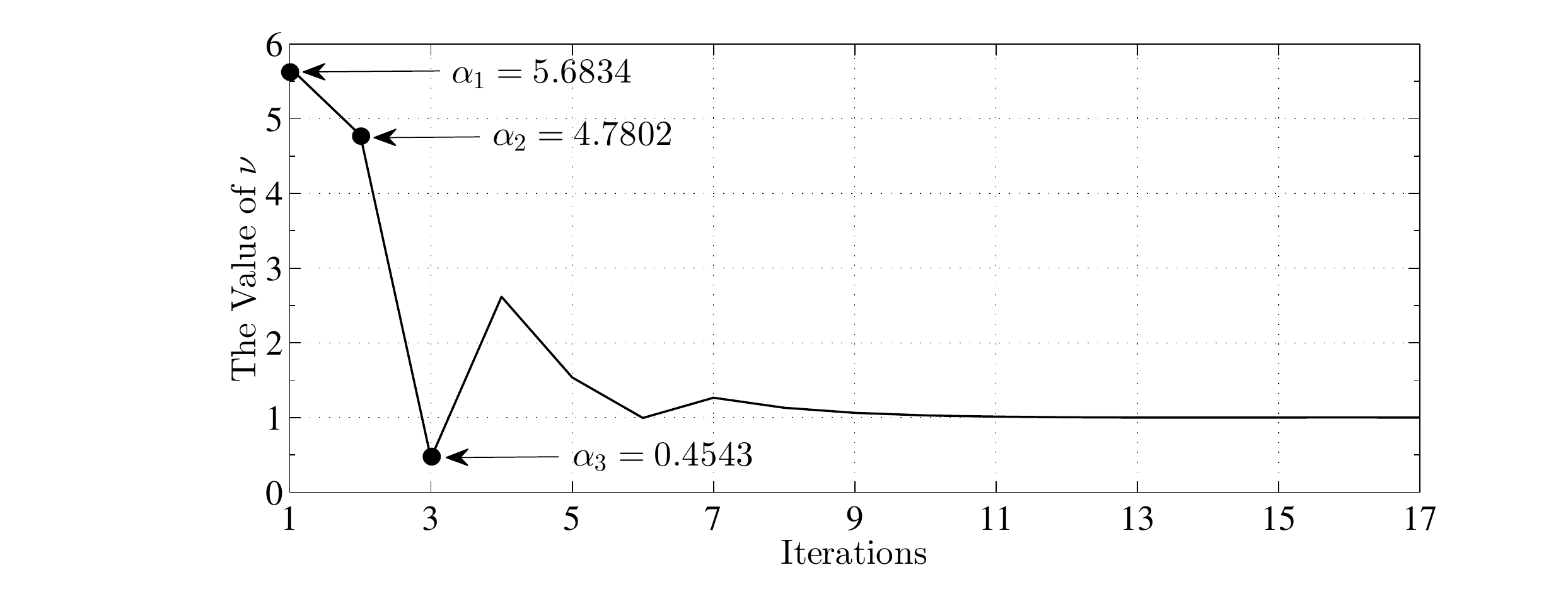}
  \label{Fig.convcoop}}}
\hfil
 \subfigure[Convergence of the auction based distributed algorithm.]
  {\scalebox{0.35}{\includegraphics {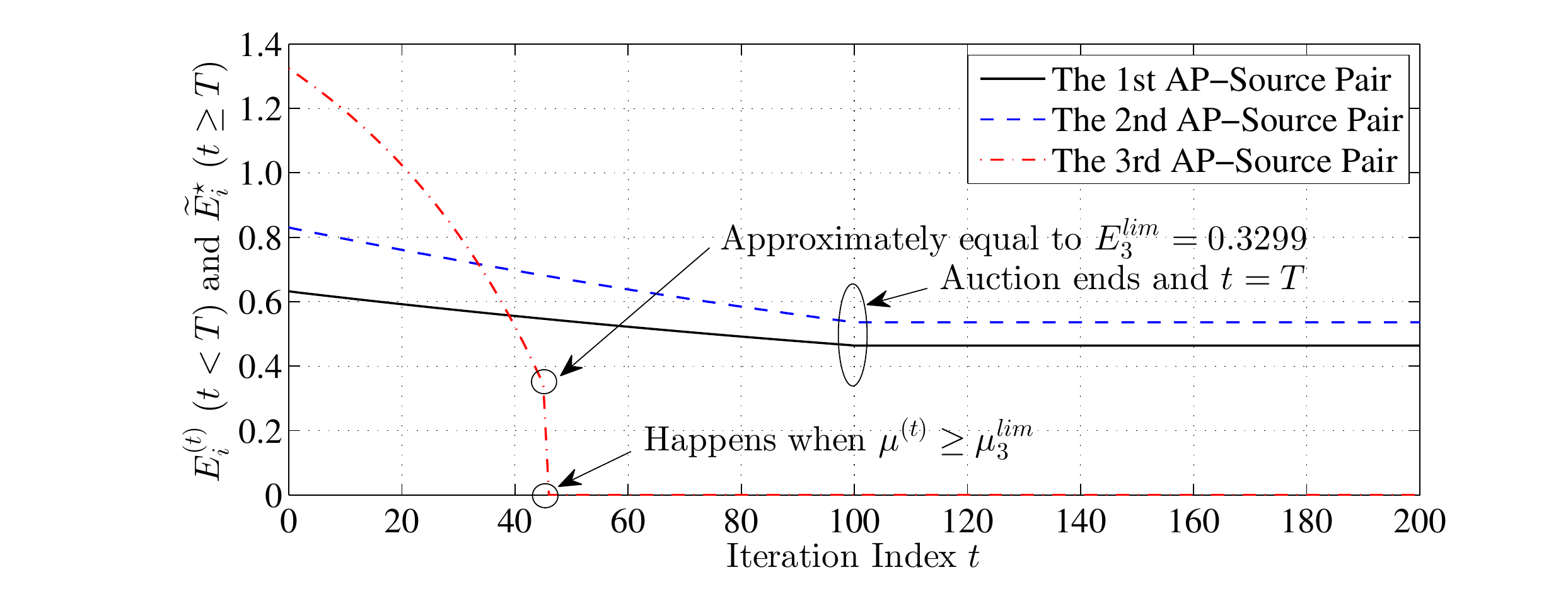}
\label{Fig.convnoncoop}}}
\caption{The convergence properties of both the proposed algorithms.}
\label{fig3}
\end{figure}

\begin{figure*}[t]
\centering
 \subfigure[The optimal allocated energy to each AP-source pair versus $E_b^{tot}$ in the case of $N = 3$ for both the proposed scenarios.]
  {\scalebox{0.41}{\includegraphics {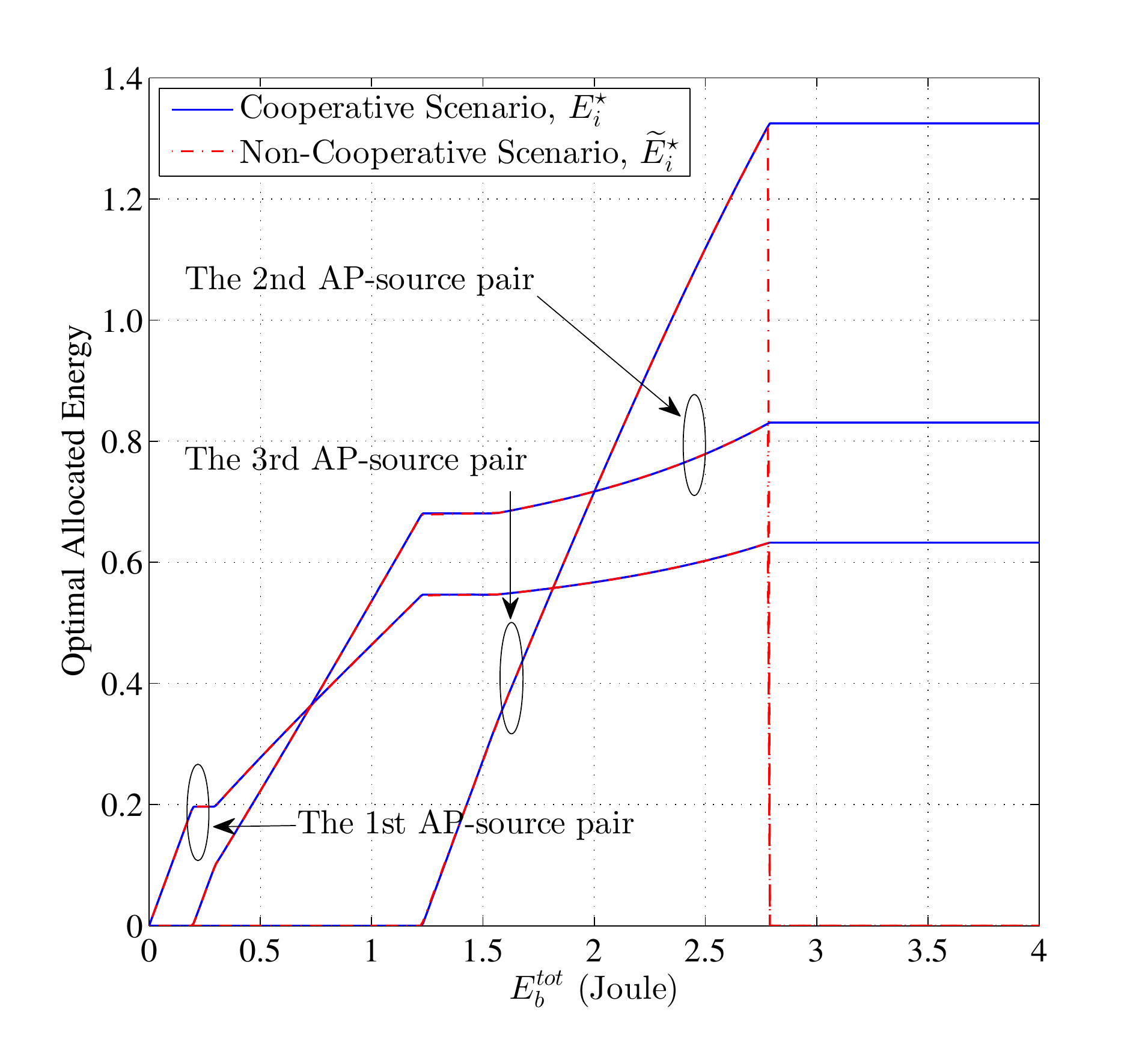}
  \label{fig.4a}}}
\hfil
 \subfigure[The optimal DL WET time to each source versus $E_b^{tot}$ in the case of $N = 3$ for both the proposed scenarios.]
  {\scalebox{0.41}{\includegraphics {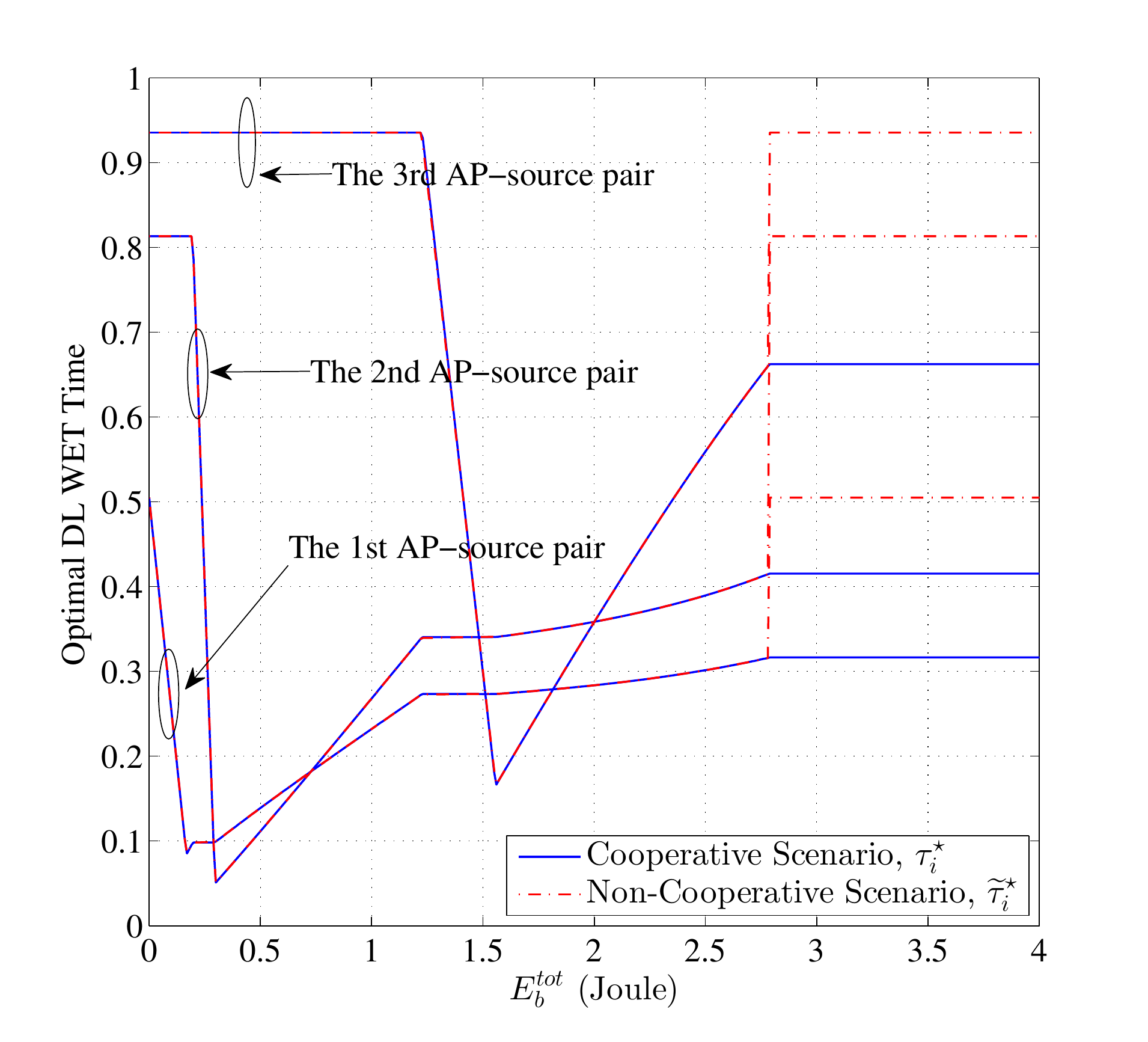}
\label{fig.4b}}}
\caption{The optimal allocated energy and optimal DL WET time for each AP-source pair versus $E_b^{tot}$ in the case of $N = 3$ for both the proposed cooperative and non-cooperative scenarios.}
\label{fig4}
\end{figure*}

\begin{figure*}[t]
\centering
 \subfigure[The average allocated energy from the PB to each AP-source pair versus $E_b^{tot}$ for both the proposed scenarios.]
  {\scalebox{0.41}{\includegraphics {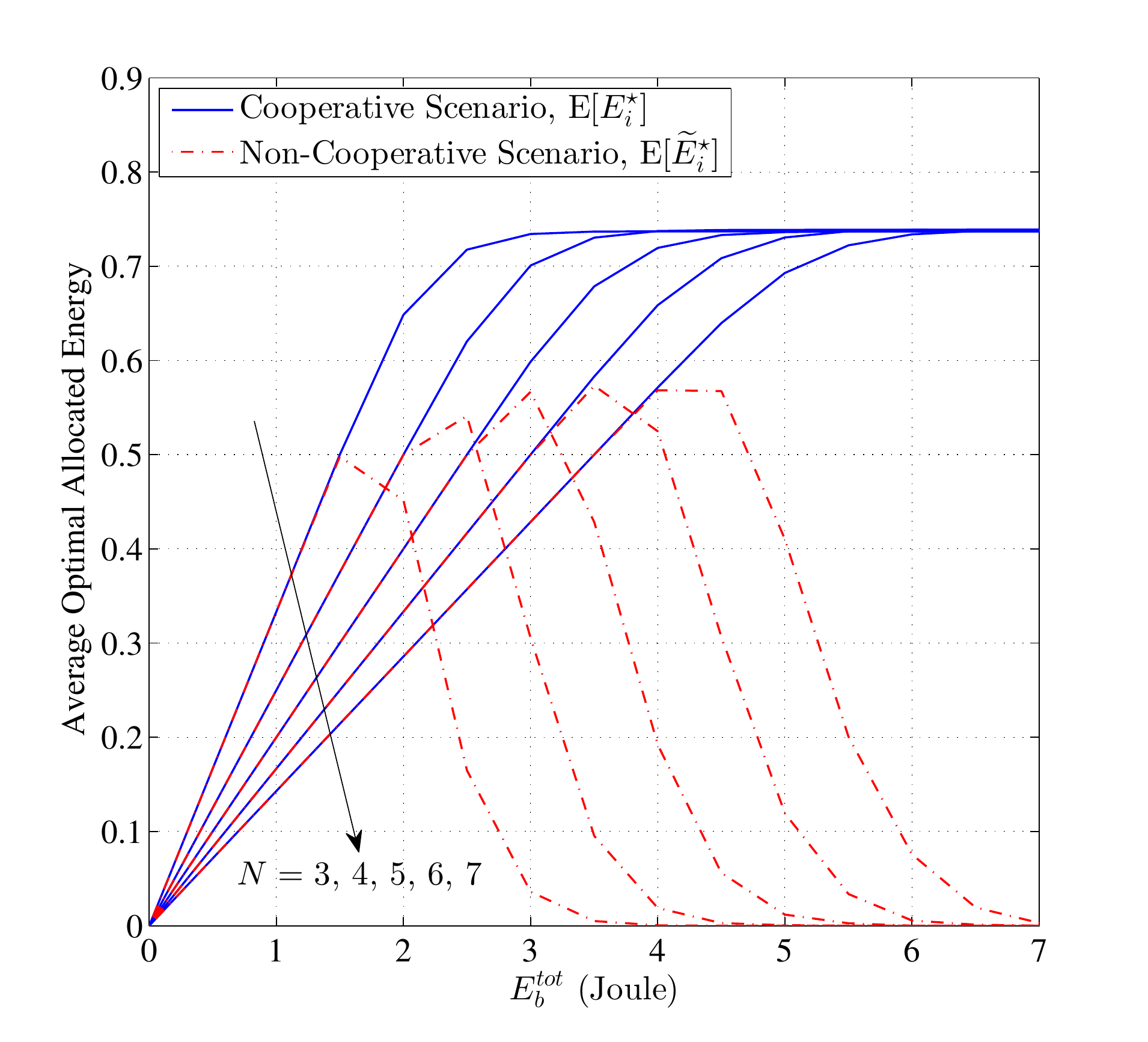}
  \label{fig.5a}}}
\hfil
 \subfigure[The average DL WET time to each source versus $E_b^{tot}$ for both the proposed scenarios.]
  {\scalebox{0.41}{\includegraphics {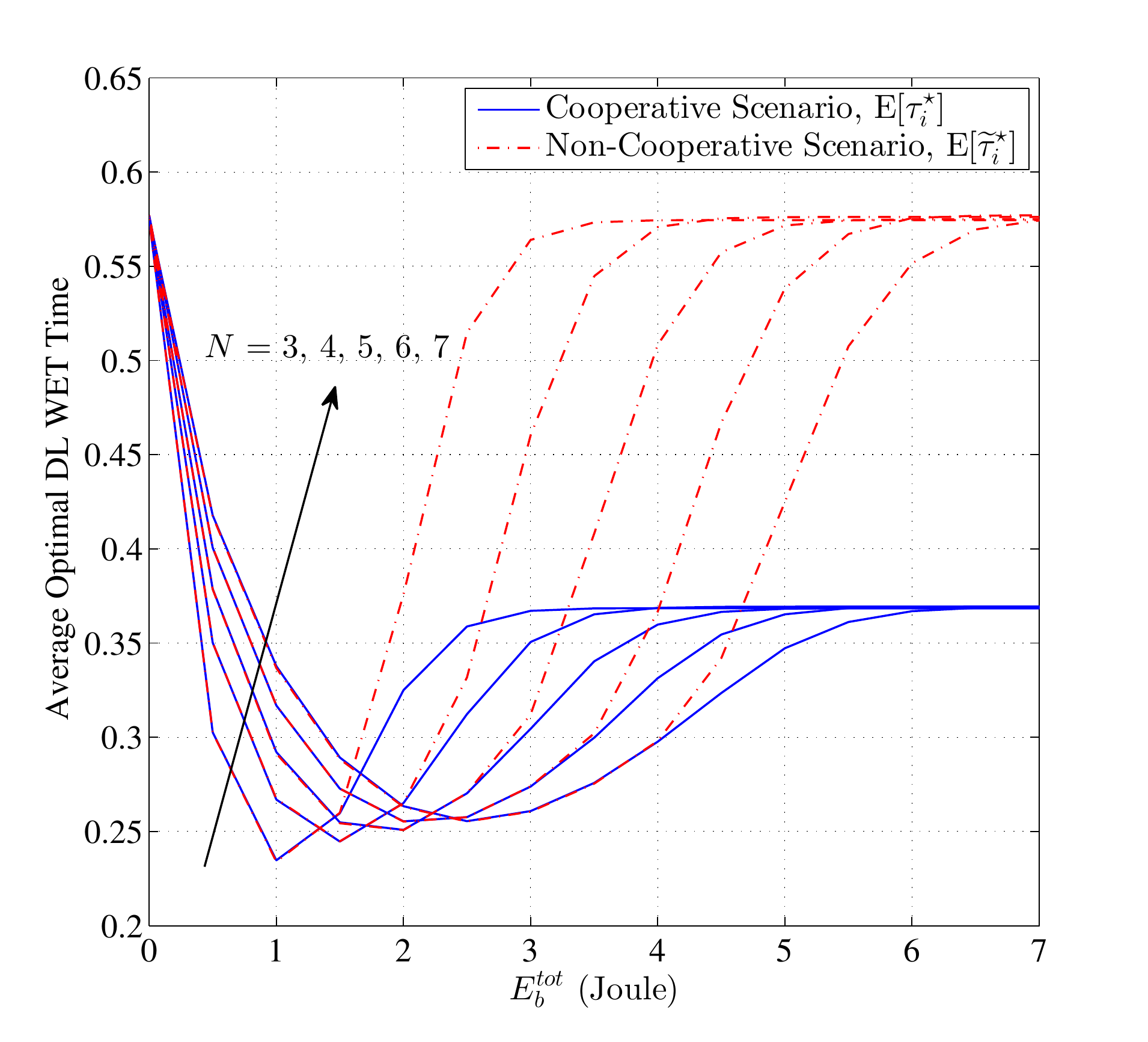}
\label{fig.5b}}}
\caption{The average allocated energy and average DL WET time for each AP-source pair versus $E_b^{tot}$ in the case of both the proposed cooperative and non-cooperative scenarios.}
\label{fig5}
\end{figure*}

\section{Numerical Results}
In this section, we present numerical results to illustrate and compare the performance of both the proposed cooperative and non-cooperative scenarios. We denote by $d_{A_iS_i}$ and $d_{PS_i}$ the distance between the $i$th source and its AP, and the distance between the $i$th source and the PB, respectively, $\forall i\in \mathcal{N}$. We assume that all the channels experience quasi-static flat Rayleigh fading and adopt a distance-dependent pass loss model such as $L_{d_{XY}} = 10^{-3} \left(d_{XY}\right)^{-\zeta}$, where $d_{XY}$ denotes $d_{A_iS_i}$ or $d_{PS_i}$ and $\zeta \in [2,5]$ is the path-loss factor. Notice that a $30$dB average signal power attenuation is assumed at a reference distance of $1$m in the above channel model~\cite{6678102}. The system parameters used in the following simulations are listed in Table \ref{tab}.

The curves in Figs. \ref{fig3}-\ref{fig4} correspond to a network setup consisting of three AP-source pairs with one randomly generated channel realization, where the channel power gains between the APs and sources and the equivalent channel power gains between the PB and the sources are
\begin{equation}
\begin{split}
\boldsymbol{G}
= \left[
                   \begin{array}{c}
                     0.0446\\
                     0.1569 \\
                     0.8628 \\
                   \end{array}
                 \right]\times 10^{-5},~
\boldsymbol{K} = \left[
                   \begin{array}{c}
                     0.1616 \\
                     0.6486 \\
                     0.4379 \\
                   \end{array}
                 \right]\times 10^{-4}.
\end{split}
\end{equation}
With these parameters, we thus can compute $\left[\alpha_1, \alpha_2, \alpha_3\right]$ = $\left[\mu^{lim}_1\right.$, $\mu_2^{lim}$, $\left.\mu_3^{lim} \right]$ = $[5.6834$, $4.7802$, $0.4543]$, $\left[E_1^{lim}\right.$, $E_2^{lim}$, $\left.E_3^{lim}\right]$ = $[0.1676$, $0.0989$, $0.3299]$ and $\left[E_1^o, E_2^o, E_3^o\right]$ = $[0.6325$, $0.8307$, $1.3247]$ based on (\ref{Eq.alpha}), (\ref{E_lim}) and (\ref{E_o}), respectively.

\begin{figure}[!t]
\centering \scalebox{0.41}{\includegraphics{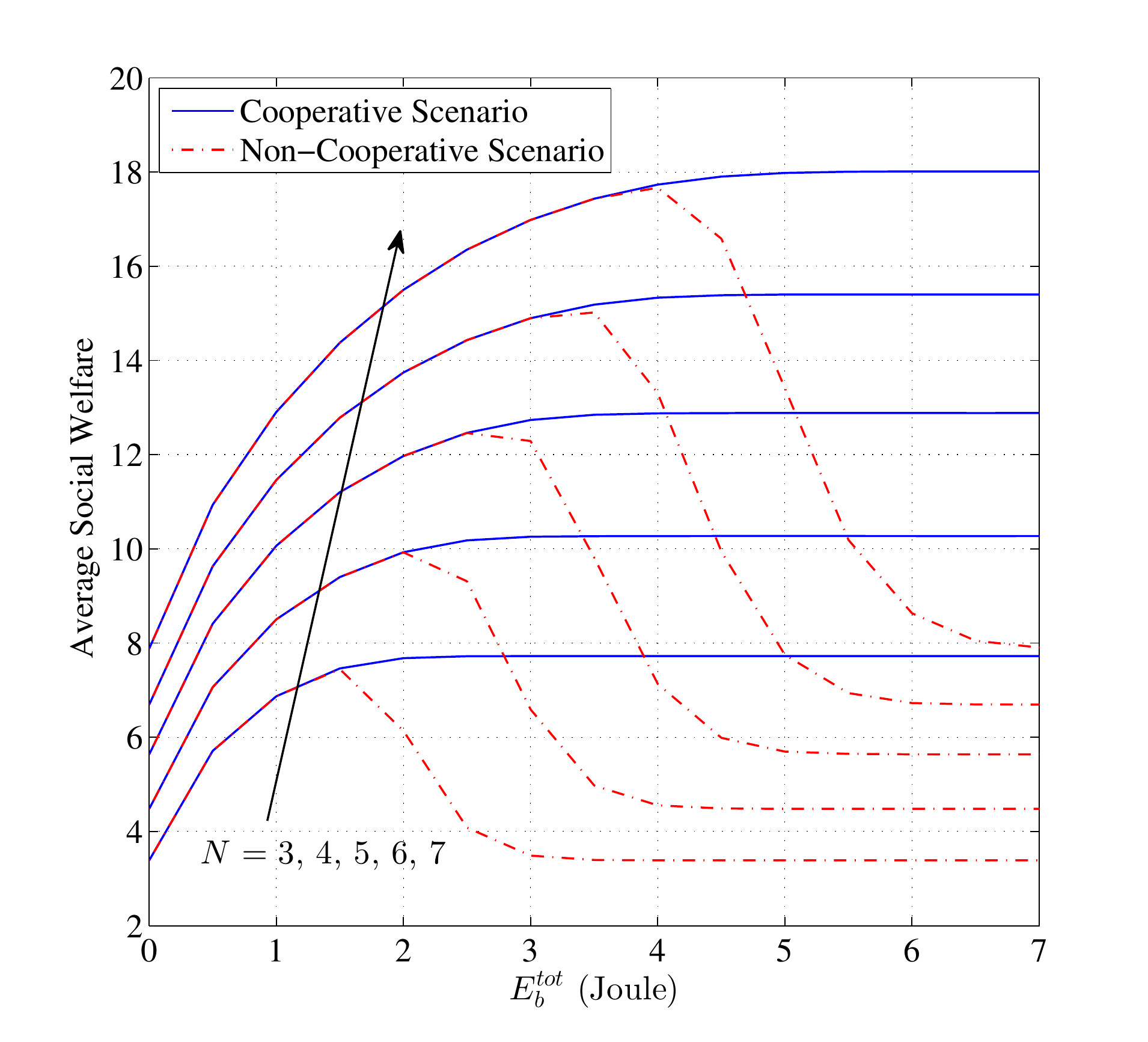}}
\caption{The average social welfare (i.e., the weighted sum throughput) of both the proposed cooperative and non-cooperative scenarios versus $E_b^{tot}$ in the case of different number of participating AP-source pairs. The social welfare in the non-cooperative scenario is actually the aggregate of the utilities of all the AP-source pairs and the PB.}
\label{fig.6}
\end{figure}

Fig. \ref{fig3} illustrates the convergence properties in the case $E_b^{tot} = 1$Joule\footnote{We consider $\mathcal{T} = 1$s to guarantee the consistency of all physical units.} for both the proposed water-filling based distributed algorithm in Algorithm \ref{Algorithm1} and the auction based distributed algorithm in Algorithm \ref{Algorithm2}. It is shown in Fig. \ref{Fig.convcoop} that the value of $\nu$ starts from $\nu = \max_{i\in \mathcal{N}}\{\alpha_i\} = \alpha_1 = 5.6834$, turns to $\nu = \alpha_2 = 4.7802$ and $\nu = \alpha_3 = 0.4543$. Then it is confirmed that $\nu$ exists between $\alpha_3$ and $\alpha_2$. With the bisection method, the proposed water-filling based distributed algorithm converges within few iterations and the desired $\nu$ is achieved. In Fig. \ref{Fig.convnoncoop}, with the increasing price $\mu^{(t)}$ (as the iteration index $t$ increases), the third AP-source pair first quits the auction when $\mu^{(t)} \geq \mu_3^{lim}$ and its last bid is approximately equal to $E_3^{lim} = 0.3299$ (when the price step $\Delta$ is sufficiently small). When the price $\mu^{(t)}$ continues to rise, the bids of the first and the second AP-source pairs will decrease till the summation of the bids is smaller than $E_b^{tot}$. Then the PB concludes the auction, sets $t = T$ and allocates the energy to each AP-source pair based on (\ref{Eq.Tclinch}), which validates the convergence of the proposed auction based distributed~algorithm.

Fig. \ref{fig4} shows the optimal allocated energy and optimal DL WET time for each AP-source pair versus $E_b^{tot}$ in the case of $N = 3$ for both the proposed cooperative and non-cooperative scenarios.
It can be observed in Fig. \ref{fig.4a} that the first AP-source pair, which has the largest $\alpha_1 = 5.6834$, is first allocated energy by the PB. Then the second and the third AP-source pairs with $\alpha_2 = 4.7802$ and $\alpha_3 = 0.4543$ are allocated energy by the PB subsequently. With the increasing of $E_b^{tot}$, the optimal allocated energy $E_i^\star$ for each AP-source pair in the cooperative scenario converges to $E_i^o$ and equals $E_i^o$ when $E_b^{tot} \geq \sum_{i\in\mathcal{N}}E_i^o$. Meanwhile, as expected, the final allocated energy $\widetilde{E}_i^\star$ for each AP-source pair in the non-cooperative scenario can efficiently match that in the cooperative scenario (i.e., $E_i^\star$), when $E_b^{tot}$ is small. But $\widetilde{E}_i^\star$ plummets to zero since the value of $E_b^{tot}$ exceeds that of $\sum_{i\in\mathcal{N}}E_i^{(0)}$, which is due to the quitting of the PB in the adopted auction mechanism.
Note that when $\mu^{(0)}$ is sufficiently small, the energy payment of each AP to the PB can be ignored in its own utility function. In this case, the non-cooperative scenario can be approximated to the cooperative scenario. Thus, the optimal bid $E_i^{(0)}$ from each AP should approximately equal to $E_i^o$, which leads to $\sum_{i\in\mathcal{N}}E_i^{(0)} \approx \sum_{i\in\mathcal{N}}E_i^{o}$. Therefore, the approach of $E_i^\star$ to $E_i^o$ and the plummeting of $\widetilde{E}_i^\star$ to zero occur almost simultaneously, as depicted in Fig. \ref{fig.4a}.

Fig. \ref{fig.4b} shows the trends of both $\tau_i^\star$ and $\widetilde{\tau}_i^\star$ versus $E_b^{tot}$. With the increasing of $E_b^{tot}$, the optimal DL WET time of the first AP-source pair first decreases and then increases while others remain the same. This is because the first AP-source pair is allocated energy first. When $E_b^{tot}$ is small, the harvested energy for the source mainly comes from its associated AP, which corresponds to the case $\tau_i > \tau_i^\prime$ (see Proposition \ref{Pro1}). But, with the assistance of the PB, the AP can shorten its DL WET time, which leads to the decrease of $\tau_i^\star$ (or $\widetilde{\tau}_i^\star$). When $E_b^{tot}$ continually increases, the source can harvest more energy from the PB, which corresponds to the case that the DL WET time is dominated by the PB, i.e., $\tau_i = \tau_i^\prime$ (see Proposition \ref{Pro1}). In this case, the value of $\tau_i^\star$ (or $\widetilde{\tau}_i^\star$) grows as the value of $E_b^{tot}$ increases. Moreover, we can observe the same tendency of the second and the third AP-source pairs when they are allocated energy. When $E_i^{tot} \geq \sum_{i\in\mathcal{N}}E_i^o$, $\tau_i^\star$ of each AP-source pair in the cooperative scenario becomes constant as $E_i^\star = E_i^o$, almost at the same time, $\widetilde{\tau}_i^\star$ for each AP-source pair in the non-cooperative scenario equals that in the case each source only harvests energy from its own AP, which is caused by the quitting of the PB. 

Now we show the average performance of both the proposed scenarios and each curve hereafter is averaged over $10^4$ randomly generated channel realizations. Fig. \ref{fig5} demonstrates the average allocated energy and average DL WET time for each AP-source pair versus $E_b^{tot}$ in both the proposed cooperative and non-cooperative scenarios. $\mathbb{E}\left[E_i^\star\right]$, $\mathbb{E}[\widetilde{E}_i^\star ]$, $\mathbb{E}\left[\tau_i^\star\right]$ and $\mathbb{E}\left[\widetilde{\tau}_i^\star\right]$ are averaged for each AP-source pair and averaged over $10^4$ channel realizations. It can be observed in Fig.~\ref{fig.5a} that the optimal allocated energy in the cooperative scenario $\mathbb{E}\left[E_i^\star\right]$ rises with the increasing of $E_b^{tot}$. But $\mathbb{E}\left[E_i^\star\right]$ decreases with the increment of the number of participating AP-source pairs with a given $E_b^{tot}$ because $\mathbb{E}\left[E_i^\star\right]$ is averaged for each AP-source pair. All $\mathbb{E}\left[E_i^\star\right]$s corresponding to different $N$ converge to a constant value with the increasing of $E_b^{tot}$, because the larger $E_b^{tot}$ is, the larger is the probability of $E_b^{tot} \geq \sum_{i\in\mathcal{N}} E_i^o$ in each simulation block with one channel realization.
For the non-cooperative scenario, $\mathbb{E}[\widetilde{E}_i^\star]$ closely matches $\mathbb{E}\left[E_i^\star\right]$ when $E_b^{tot}$ is small, but converges to zero when $E_b^{tot}$ is sufficiently large due to the quitting of the PB.
Fig. \ref{fig.5b} depicts both $\mathbb{E}\left[\tau_i^\star\right]$ and $\mathbb{E}\left[\widetilde{\tau}_i^\star\right]$ versus $E_b^{tot}$. When $E_b^{tot}$ increases, both $\mathbb{E}\left[\tau_i^\star\right]$ and $\mathbb{E}\left[\widetilde{\tau}_i^\star\right]$ decrease first and then increase, and they converge to different values for the reason shown in Fig. \ref{fig.4b}. Besides, when the number of participating AP-source pairs rises, the turning point of the corresponding curve appears later. The reason is that the more AP-source pairs, the less energy allocated to each source from the PB. Then, a larger value of $E^{tot}_b$ is needed to make the DL WET time at each source dominated by the PB.

Fig. \ref{fig.6} presents the average social welfare of both the proposed cooperative and non-cooperative scenarios versus $E_b^{tot}$ in the case of different number of participating AP-source pairs. The social welfare in the non-cooperative scenario is assumed to be the aggregate of the utilities of all the AP-source pairs and the PB. It can be observed that the network setup with a larger $N$ in either proposed scenario achieves a better social welfare performance due to the energy resource brought by the new joined AP-source pairs. The average social welfare of the cooperative scenario increases as $E_b^{tot}$ grows and approaches a constant value when $E_b^{tot}$ is large enough. This is because when $E_b^{tot} \geq \sum_{i\in\mathcal{N}}E_i^o$, as shown in Fig. \ref{fig.4a}, the allocated energy to each source will be a constant value and independent of $E_b^{tot}$, which leads to the saturation of the social welfare performance. In the non-cooperative scenario, as $E_b^{tot}$ increases, the average social welfare performance closely matches that in the cooperative scenario when $E_b^{tot}$ is small, but then deteriorates due to the quitting of the PB when $E_b^{tot} \geq \sum_{i\in\mathcal{N}}E_i^{(0)}$. Notice that when $E_b^{tot} =0$, i.e., without the PB, the average social welfare performance is always the~worst.

\section{Conclusions}

This paper investigated the joint time and energy allocation of a power beacon (PB)-assisted wireless-powered communication network (WPCN) with multiple access point (AP)-source pairs and a PB.
We consider both cooperative and non-cooperative scenarios corresponding to the situations whether the PB provides wireless charing service to each AP-source pair for free or not.
Moreover, the social welfare was maximized in the proposed cooperative scenario and the respective utility of each AP and the PB was maximized in the non-cooperative scenario. The numerical results validated the convergence of both the proposed water-filling based and auction based distributed algorithms. It was demonstrated that the average social welfare of the cooperative scenario improves as either the number of participating AP-source pairs or the total energy of the PB increases, but saturates when the total energy of the PB is sufficiently large. Moreover, the average social welfare performance of the non-cooperative scenario closely matches that in the cooperative scenario when the total energy of the PB is small, but deteriorates when the total energy of the PB is sufficiently large, which is caused by the quitting of the PB in the adopted auction mechanism.

\section*{Appendix A\\Proof of Proposition \ref{Pro1}}
To proceed, we first derive the second-order derivative of $\mathcal{S}_i\left(\tau_i\right)$ with respect to~$\tau_i$. After some simplifications, we have
\begin{equation}
\begin{split}
\frac{\partial^2 \mathcal{S}_i\left(\tau_i \right)}{\partial {\tau_i}^2} = - \frac{ \lambda_i W \left(A_i + B_i\right)^2  }{\left(1-\tau_i\right)\left(1-\tau_i + A_i \tau_i + B_i \right)^2 \ln2}.
\end{split}
\end{equation}
where $A_i  = \frac{G_i^2 \eta p_i }{\sigma^2}$ and $B_i = \frac{G_i \eta E_i K_i}{\sigma^2}$. Since $\frac{\partial^2 \mathcal{S}_i\left(\tau_i \right)}{\partial {\tau_i}^2} < 0$, $\mathcal{S}_i\left(\tau_i\right)$ is a concave function of $\tau_i$. Thus, the optimal solution can be obtained by setting the first-order derivative of $S_i(\tau_i)$ equal to zero and comparing the obtained stationary points with the constraints. That is, $\frac{\partial \mathcal{S}_i\left(\tau_i\right)}{\partial \tau_i } = 0$. After some algebraic manipulations, we have
\begin{equation}\label{A.2}
\begin{split}
{z_i}\ln\left({z_i}\right) - {z_i} + 1 = A_i,
\end{split}
\end{equation}
where $z_i = 1+ \frac{ A_i \tau_i  + B_i }{1-\tau_i}$. Note that $z_i > 1$ as $A_i > 0$ and $0 < \tau_i < 1$.
With reference to \cite{6678102}, $f\left(z_i\right) = {z_i}\ln\left({z_i}\right) - {z_i} + 1$ is strictly increasing when $z_i > 1$ and $f\left(1\right) = 0$. We then can deduce that there exists a unique solution, denoted by $z_i^\dag >1$, to the equality $f(z_i) = A_i$. After some manipulations, we get
\begin{equation}
\begin{split}
{z_i^\dag }= \exp\left(\mathcal{W}\left( \frac{A_i  - 1}{\exp(1)}\right) +1\right),
\end{split}
\end{equation}
in which $\mathcal{W}\left(x\right)$ is the Lambert $\mathcal{W}$ function that is the solution to the equality $x = \mathcal{W}\exp(\mathcal{W})$. Let $\tau_i^\dag$ denote the optimal solution to the equation (\ref{A.2}). Then, we have $1+ \frac{ A_i \tau_i^\dag  + B_i }{1-\tau_i^\dag} = z_i^\dag$. Rearranging this equality, we get
\begin{equation}\label{A.3}
\begin{split}
\tau_i^\dag = \frac{ z_i^\dag - 1 -B_i }{{z_i^\dag - 1 +  A_i}}.
\end{split}
\end{equation}

Recall the constraints on $\tau_i$ in the optimization problem (\ref{Pro.tau}) that $ 0 < \tau_i < 1$ and $\tau_i \geq \frac{E_i}{p_b}$, we have
\begin{equation}\label{A.5}
\tau_i\left(E_i\right) =\max\left\{ \tau_i^\dag, \frac{E_i}{p_b}\right\},
\end{equation}
since it is easy to check that $\tau_i^\dag<1$ and $\tau_i^\dag > 0$ when $E_i = 0$. To further expand (\ref{A.5}),  we compare the two terms $\tau_i^\dag$ and $E_i/p_b$ and obtain that $\tau_i^\dag < E_i/p_b$ only if
\begin{equation}\label{A.6}
\begin{split}
E_i > \frac{p_b \left(z_i^\dag -1\right) \sigma^2 }{\left(z_i^\dag -1\right)\sigma^2 + G_i\eta \left(p_i G_i + p_b K_i\right)},
\end{split}
\end{equation}
which is obtained by substituting the expression of $A_i$ and $B_i$ into (\ref{A.3}). Therefore, the optimal solution ${\tau}_i\left(E_i\right)$ to the problem (\ref{Pro.tau}) can be further expressed as
\begin{equation}\label{A.4}
\tau_i\left(E_i\right) =
\begin{cases}
   \tau_i^\dag, &\mbox{if ~$0 \leq E_{i}\leq E_i^{lim}$},\\
   \frac{E_{i}}{p_b}, &\mbox{if ~$ E_i^{lim}< E_i < p_b$},
\end{cases}
\end{equation}
where $E_i^{lim}$ is defined as the right-hand side of (\ref{A.6}). This completes the proof.

\section*{Appendix B\\Proof of Lemma \ref{Lem2} and Proposition \ref{Pro2}}
When $0 \leq E_i \leq E_i^{lim}$, by replacing $\tau_i$ in (\ref{Eq.tauE}) with $\tau_i\left(E_i\right)$, with careful simplification, we~have
\begin{equation}
\begin{split}
\mathcal{S}_i\left( E_i \right) = & \lambda_i W \frac{ G_i\eta \left( p_i G_i + E_i K_i\right) }{{\left(z_i^\dag - 1\right) \sigma^2 +  G_i^2 \eta p_i}} \log_2\left(z_i^\dag\right)\\
 = &\lambda_i W \frac{ G_i\eta \left( p_i G_i + E_i K_i\right) }{{\left(z_i^\dag - 1\right) \sigma^2 +  G_i^2 \eta p_i}} \frac{\left(z_i^\dag - 1\right) \sigma^2 +  G_i^2 \eta p_i}{z_i^\dag \sigma^2 \ln 2}\\
 = & \frac{\lambda_i W G_i \eta \left( p_i G_i + E_i K_i\right)}{z_i^\dag \sigma^2 \ln 2}.
\end{split}
\end{equation}
where the second equality is based on that $z_i^\dag \ln\left(z_i^\dag\right)- z_i^\dag +1 = \frac{G_i^2 \eta p_i}{\sigma^2}$ (please refer to the proof in Appendix A).
Thus, the gradient of $\mathcal{S}_i \left( E_i \right)$ is a constant, denoted by $\alpha_i$, given by
\begin{equation}
\begin{split}
\alpha_i =\frac{\partial \mathcal{S}_i\left( E_i \right)}{\partial E_i} = \frac{ \lambda_i W G_i\eta K_i}{z_i^\dag \sigma^2 \ln 2} > 0.
\end{split}
\end{equation}

When $E_i^{lim} < E_i < p_b$, by inserting the expression of $\tau_i$ into (\ref{Eq.tauE}), we have
\begin{equation}
\begin{split}
\mathcal{S}_i\left( E_i \right) = \lambda_i  W \left(1-\frac{E_i}{p_b}\right) \log_2 \left(  1+  \frac{X_i E_i   }{ p_b- E_i}   \right),
\end{split}
\end{equation}
where $X_i \triangleq \frac{G_i\eta\left(p_i G_i +p_b K_i\right)}{\sigma^2}$. By deriving the first-order derivative of $\mathcal{S}_i \left( E_i \right)$, we have the gradient of $\mathcal{S}_i \left( E_i \right)$ when $E_i^{lim} < E_i < p_b$, denoted by $\beta_i\left(E_i\right)$, as
\begin{equation}\label{B.1}
\begin{split}
\beta_i\left(E_i\right) = \frac{\partial \mathcal{S}_i\left( E_i \right)}{\partial E_i} = &-\frac{\lambda_i W}{p_b}\log_2\left(  1+  \frac{X_i E_i   }{ p_b- E_i}   \right)\\
 &+ \frac{\lambda_i W X_i }{\left(p_b - E_i + X_i E_i\right)\ln2}.
\end{split}
\end{equation}
Moreover, we derive the second-order derivative of $\mathcal{S}_i \left( E_i \right)$ and obtain
\begin{equation}
\begin{split}
\frac{\partial^2 \mathcal{S}_i\left( E_i \right)}{\partial {E_i}^2} = - \frac{\lambda_i W X_i^2 p_b}{ \left(p_b - E_i + X_i E_i \right)^2 \left(p_b - E_i\right)\ln2} < 0.
\end{split}
\end{equation}
Thus, when $E_i^{lim} < E_i < p_b$, $\beta_i\left(E_i\right)$ is strictly~decreasing.

Furthermore, by inserting the expression of $E_i^{lim}$, given by (\ref{E_lim}), into $\beta_i\left(E_i\right)$, with simplification, we have
\begin{equation}
\begin{split}
\beta_{i}\left(E_i^{lim}\right) &= \frac{\lambda_i W}{z_i^\dag p_b\ln2} \left(X_i - \left(z_i^\dag \ln\left(z_i^\dag\right)- z_i^\dag +1\right)\right)\\
 &=  \frac{\lambda_i W}{z_i^\dag p_b\ln2} \left(\frac{G_i\eta\left(p_i G_i +p_b K_i\right)}{\sigma^2} - \frac{G_i^2 \eta p_i}{\sigma^2}\right)\\
 &= \alpha_i.
\end{split}
\end{equation}
Hence, on $0 \leq E_i < p_b$, $\mathcal{S}_i \left( E_i \right)$ is differentiable and the gradient of $\mathcal{S}_i\left(E_i\right)$ is continuous.

\section*{Appendix C\\Proof of Proposition \ref{Pro4}}

According to Proposition \ref{Pro3}, the Karush-Kuhn-Tucker (KKT) conditions are both necessary and sufficient for the optimality of the problem (\ref{Eq.SWO2}). To proceed, we first discuss the solution of a problem without the condition $\boldsymbol{0} \preceq \boldsymbol{E} \preceq \boldsymbol{E^o}$. By removing $\boldsymbol{0} \preceq \boldsymbol{E} \preceq \boldsymbol{E^o}$, we have the Lagrangian as
\begin{equation}
\begin{split}
\mathcal{L} = \sum_{i\in\mathcal{N}}\mathcal{S}_i\left(E_i\right) - \nu\left( \sum_{i\in\mathcal{N}} E_i - E_b^{tot}\right),
\end{split}
\end{equation}
Taking the stationarity condition, we have
\begin{equation}\label{C.1}
\begin{split}
\nabla \mathcal{S}_i\left(E_i\right) - \nu = 0,
\end{split}
\end{equation}
and the KKT conditions are
\begin{subequations}\label{C.3}
\begin{align}
&\nu\left( \sum_{i\in\mathcal{N}} E_i - E_b^{tot}\right) = 0,\\
& \nu \geq 0.
\end{align}
\end{subequations}

Recall that the gradient of $\mathcal{S}_i\left(E_i\right)$ is subject to that $0\leq \nabla \mathcal{S}_i\left(E_i\right) \leq \alpha_i$ when $0 \leq E_i \leq E_i^o$, on that basis, we begin to discuss the following cases.

\textbf{Case~1:} When $E_i^{lim} < E_i \leq E_i^o$, $\nabla \mathcal{S}_i\left(E_i\right) = \beta_i\left(E_i\right)$, $\beta_i\left(E_i\right)$ is strictly decreasing and $0 \leq \beta_i\left(E_i\right) <\alpha_i $. Therefore, given $0 \leq \nu < \alpha_i$, there always exists a unique $E_i^\star$ satisfying (\ref{C.1}). By solving that $\beta\left(E_i^\star\right) = \nu$ with (\ref{Eq.beta}), we have
\begin{equation}\label{C.2}
\begin{split}
z_i \ln\left(z_i\right) + \left(\frac{\nu p_b\ln2}{\lambda_i W} -1\right)z_i + 1 = X_i,
\end{split}
\end{equation}
where $X_i =\frac{G_i\eta\left(p_i G_i +p_b K_i\right)}{\sigma^2}$, $z_i = 1+ \frac{X_i E_i^\star}{p_b - E_i^\star}$ and $z_i > 1$ as $X_i > 0$ and $E_i^{lim} < E_i^\star \leq E_i^o$. We denote by $z_i^\S > 1$ the solution of (\ref{C.2}). Then, we denote $E_i^\star = \gamma_i\left(\nu\right)$ and~obtain $\gamma_i\left(\nu\right)$ by solving $z_i^\S = 1+ \frac{X_i E_i^\star}{p_b - E_i^\star}$ as
\begin{equation}
\begin{split}
\gamma_i\left(\nu\right)= \frac{p_b\left(z_i^\S -1\right)\sigma^2}{\left(z_i^\S - 1\right)\sigma^2 +G_i\eta\left(p_i G_i +p_b K_i\right)}.
\end{split}
\end{equation}

We then prove that $z_i^\S$ is unique. We first define a function that $g\left(z_i\right) = z_i \ln \left(z_i\right) + \left(Y_i -1\right) z_i + 1$, which is the left-hand side of (\ref{C.2}) and $Y_i = \frac{\nu p_b\ln2}{\lambda_i W}$. By deriving the first-order derivative of $g\left(z_i\right)$, we have
\begin{equation}
\begin{split}
\frac{\partial g\left(z_i\right)}{\partial z_i} = \ln z_i + Y_i.
\end{split}
\end{equation}
Therefore, as $Y_i \geq 0$, $g\left(z_i\right)$ is monotonically increasing when $z_i > 1$. When $z_i = 1$, we have that $g\left(1\right) = Y_i$. Consequently, there exists a unique solution $z_i^\S > 1$ of the function $g\left(z_i\right) = X_i$, if $X_i > Y_i$. By comparing $X_i$ and $Y_i$, we have
\begin{equation}
\begin{split}
X_i - Y_i  = & \frac{G_i\eta \left(p_i G_i + {p_b} K_i\right)}{\sigma^2} - \frac{\nu p_b \ln 2}{\lambda_i W}\\
>& \frac{G_i\eta {p_b} K_i}{\sigma^2} - \frac{\nu p_b \ln 2}{\lambda_i W}\\
> &\frac{p_b\ln2}{\lambda_i W} \left( \frac{\lambda_i W  G_i \eta K_i}{\sigma^2 \ln 2} - \alpha_i\right) > 0.
\end{split}
\end{equation}
Hence, $z_i^\S > 1$ is the unique solution of (\ref{C.2}).

Furthermore, as $\beta_i\left(E_i\right)$ is strictly decreasing, $\gamma_i\left(\nu\right)$ is strictly decreasing as well. Because of the continuous property of $\nabla \mathcal{S}_i\left(E_i\right)$, we have $\gamma_i\left(0\right) = E_i^o$ and $\gamma_i\left(\alpha_i\right) = E_i^{lim}$.

\textbf{Case 2:} When $0 < E_i \leq E_i^{lim}$, $\nabla \mathcal{S}_i\left(E_i\right) = \alpha_i$. Thus, given $\nu = \alpha_i$, any value of $E_i$ between $0$ and $E_i^{lim}$ could satisfy (\ref{C.1}). However, by considering the KKT conditions, we have that $E_i^\star = E_b^{tot} - \sum_{j \in\mathcal{N}\backslash\{i\}} E_j^\star$.

The remaining case of $\nu > \alpha_i$, leads to $E_i^\star = 0$. In summary, we have proved Proposition \ref{Pro4}, where $\nu$ is chosen to meet the total energy constraint that $\sum_{i\in \mathcal{N}} E_i^\star = E_b^{tot}$.

\section*{Appendix D\\Proof of Proposition \ref{Pro5}}

The problem (\ref{AP_OP}) can be solved by following a similar procedure for problem (\ref{Eq.SWO}). We first apply the optimal relationship between $\tau_i$ and $E_i$ and simply problem (\ref{AP_OP}) to the following one
\begin{equation}
\begin{split}
\max_{E_i}~ \mathcal{U}_i\left(E_i, \mu^{(t)} \right), ~\textrm{s.t.}~ 0 \leq E_i <p_b,
\end{split}
\end{equation}
where
\begin{equation}\label{D.1}
\mathcal{U}_i\left( E_i, \mu^{(t)} \right)=\mathcal{S}_i\left(E_i\right) - \mu^{(t)} E_i,
\end{equation}
with $\mathcal{S}_i\left(E_i\right)$ defined in (\ref{Eq.S(E)}).

Based on the gradient of $\mathcal{S}_i\left(E_i\right)$ shown in (\ref{grad}), we can readily obtain the gradient of $\mathcal{U}_i\left( E_i, \mu^{(t)} \right)$ with respect to $E_i$, denoted by $\nabla\mathcal{U}_i\left( E_i \right)$, as follows.
\begin{equation}
\nabla\mathcal{U}_i\left( E_i \right)=
\begin{cases}
    \alpha_i - \mu^{(t)},  &\mbox{if ~$ 0 \leq E_{i} \leq E_{i}^{lim}$},\\
    \beta_i\left(E_i\right) - \mu^{(t)}, &\mbox{if ~$ E_i^{lim} < E_{i} < p_b$}.
\end{cases}
\end{equation}

As $\beta_i\left(E_i^{lim}\right) = \alpha_i$ and $\beta_i\left(E_i\right)$ is strictly decreasing (refer to Appendix B), $\nabla \mathcal{U}_i\left( E_i \right)$ is continuous on $0\leq E_i < p_b$, remains constant when $0 \leq E_i \leq E_i^{lim}$ and strictly decreasing when~$E_i^{lim}< E_i < p_b$. Then, we can infer that when $\mu^{(t)} \geq \alpha_i$, $\nabla \mathcal{U}_i\left( E_i \right) \leq 0$, which results in that $E_i^{(t)} = 0$. When $\mu^{(t)} < \alpha_i$, as $\mu^{(t)} > 0$ and according to Proposition \ref{Pro3}, we have that $E_i^{(t)}$ is the unique solution of $\beta_i\left(E_i\right) - \mu^{(t)} = 0$ and $E_i^{lim}< E_i^{(t)} \leq E_i^o < p_b$. According to the analysis in Appendix C, we have $E_i^{(t)} = \gamma_i\left(\mu^{(t)}\right)$. In a summary, we have
\begin{equation}\label{D.2}
E_i^{(t)}=
\begin{cases}
   0,  &\mbox{if ~$\mu^{(t)} \geq \mu^{lim}_i$},\\
   \gamma_i\left(\mu^{(t)}\right), &\mbox{if ~$\mu^{(t)} < \mu^{lim}_i$},
\end{cases}
\end{equation}
where $\mu_i^{lim} = \alpha_i$ and it is defined as the maximum acceptable price for the $i$th AP.

With the optimal $E_i^{(t)}$, the $i$th AP thus can decide the optimal $\tau_i^{(t)}$. The utility function of the $i$th AP with the given $E_i^{(t)}$ and $\mu^{(t)}$ can be reduced to
\begin{equation}
\begin{split}
\mathcal{U}_i\left(\tau_i \right) = & \lambda_i (1-\tau_i) W  \log_2 \left(  1+ \frac{G_i \eta  \left( \tau_i p_i G_i + E_i^{(t)} K_i \right)}{\left(1-\tau_i\right)\sigma^2}    \right)\\
& - \mu^{(t)} E_i^{(t)}.
\end{split}
\end{equation}
Based on the analysis in Appendix A, we can directly obtain that
\begin{equation}\label{D.3}
\tau_i^{(t)} =
\begin{cases}
   \tau_i^\dag, &\mbox{if ~$0 \leq E_{i}^{(t)}\leq E_i^{lim}$},\\
   \frac{E_i^{(t)}}{p_b}, &\mbox{if ~$ E_i^{lim}< E_i^{(t)} < p_b$}.
\end{cases}
\end{equation}

Note that when $\mu^{(t)} < \mu_i^{lim}$, $\gamma_i\left(\mu^{(t)}\right) > \gamma_i\left(\mu_i^{lim}\right)$ and $\gamma_i\left(\mu_i^{lim}\right) = E_i^{lim}$. We now can express the optimal solution to problem (\ref{AP_OP}) in a compact form given by
\begin{equation}
\left(\tau_i^{(t)}, E_i^{(t)}\right)=
\begin{cases}
   \left(\frac{ \left(z_i^\dag - 1\right) \sigma^2}{\left(z_i^\dag - 1\right) \sigma^2 +  G_i^2 \eta p_i}, 0 \right), &\mbox{if ~$\mu^{(t)} \geq \mu^{lim}_i$},\\
   \left( \frac{\gamma_i\left(\mu^{(t)}\right)}{p_b}, \gamma_i\left(\mu^{(t)}\right)\right), &\mbox{if ~$\mu^{(t)} < \mu^{lim}_i$}.
\end{cases}
\end{equation}
This completes the proof.

\section*{Acknowledgement}
The authors would like to thank the editor and anonymous reviewers for their constructive comments and suggestions, which help us improve the quality of the paper.

\bibliographystyle{IEEEtran}
\bibliography{energy_harvesting_endnote}

\end{document}